\documentclass[letterpaper,11pt]{article}
\usepackage{jheppub} 
\usepackage{mystuff}
\usetikzlibrary{external}
\usepackage{lineno}
\newcommand{\Diff}{\text{Diff}}

\definecolor{amethyst}{rgb}{0.54, 0.17, 0.89}
\definecolor{coral}{rgb}{1.0, 0.3, 0.4}

\arxivnumber{2312.08434} 

\author[a,b,c]{Vijay Balasubramanian,}
\author[a]{and Charlie Cummings}
\affiliation[a]{David Rittenhouse Laboratory, University of Pennsylvania, \\ 209 S.33rd Street, Philadelphia, PA 19104, USA}
\affiliation[b]{Santa Fe Institute, \\1399 Hyde Park Road, Santa Fe, NM 87501, USA}
\affiliation[c]{ Theoretische Natuurkunde, Vrije Universiteit Brussel, \\Pleinlaan 2, B-1050 Brussels, Belgium}

\emailAdd{vijay@physics.upenn.edu}
\emailAdd{charlie5@sas.upenn.edu}

\newcommand{\ecs}{\mathfrak{ecs}}

\usepackage{graphicx}
\usepackage{dcolumn}
\usepackage{bm}
\usepackage{xcolor}
\usepackage{mystuff}
\newcommand{\diff}{\mathfrak{diff}}
\pgfplotsset{compat=1.18} 

\begin{document}

\title{
    The entropy of finite gravitating regions
}

\date{\today}

\abstract{
    We develop a formalism for calculating the entanglement entropy of an arbitrary spatial region of a gravitating spacetime at a moment of time symmetry. The crucial ingredient is a path integral over embeddings of the region into the overall spacetime, interpretable as a sum over the edge modes associated with the region. We argue that this entanglement entropy can be interpreted, to leading order in Newton's constant, as the minimal generalized entropy among all regions that enclose it. This suggests a notion of ``terrestrial holography'' where regions of space can encode larger ones, in contrast to the standard form of holography, in which degrees of freedom on the celestial sphere at the boundary of the universe encode the interior.
}

\maketitle 

\section{Introduction}

Gravity has a deep and mysterious connection to thermodynamics: the entropy of some gravitating systems is geometrized as the area of a distinguished surface in spacetime.  For example, black holes appear to carry a coarse-grained entropy of $S = A/4 G_N$ where $A$ is the area of the horizon and $G_N$ is Newton's constant \cite{Bekenstein:1973ur,hawkingBH}.   Cosmological \cite{Gibbons:1977mu} and Rindler \cite{Jacobson_2003,Unruh:1976db} horizons associated to accelerating observers also obey a similar relation, although the statistical interpretation of these entropies remains obscure. Associating an entropy to causal horizons can even be used to \emph{derive} the Einstein equations \cite{Jacobson_1995}.  In addition, in the context of the holographic AdS/CFT correspondence, certain extremal surfaces measure entanglement entropy between the quantum degrees of freedom associated to opposite sides of the surface \cite{Ryu_2006,Faulkner:2013ana,Hubeny:2007xt,Engelhardt:2014gca}.  
Because the thermodynamic behavior of gravity seems to hold regardless of the cosmological constant, one can ask whether there is also a relation between entropy and area for arbitrary subregions of a gravitating spacetime. Indeed, Bousso and Penington recently used geometric arguments and consistency with general constraints on quantum information to propose such a relation \cite{Bousso:2022hlz,Bousso:2023sya}.

In this paper, we investigate the entropic properties of a compactly supported gravitating region. We propose a modification of the gravitational path integral which prepares a reduced state $\rho_a$ associated with a finite subregion $a$ of a gravitational theory, which in particular is invariant under diffeomorphisms of that region. We use this reduced state (which we also motivate from the canonical perspective) to propose an entropy formula for any finite subregion of a global spacetime, regardless of the cosmological constant. 
We then argue that this entropy formula is consistent with the proposal in \cite{Bousso:2022hlz}.

Four sections follow. In Sec.~\ref{sec:edge}, we review and construct the technical tools needed to define subregions of gravitating spacetimes. We then use these tools to define and renormalize the entropy of such a subregion. In Sec.~\ref{sec:path_int}, we propose a gravitational path integral for a finite region of space. This path integral includes an integral over all embeddings of the region into  spacetime. We then take a saddlepoint approximation over the bulk fields to compute the leading order behavior of the entropy in the small coupling ($G_N \to 0$) limit. 
In Sec.~\ref{sec:path_int_2}, we then investigate the structure of this path integral after integrating out the embedding fields before ``cutting open'' the path integral to define states and Hilbert spaces.
In Sec.~\ref{sec:subregion}, we then argue that the resulting structures may provide evidence for the proposal in \cite{Bousso:2022hlz}.
We conclude in Sec.~\ref{sec:disc} by discussing interpretations and implications of our results.

\section{The entropy of subregions in gravitating spacetimes} \label{sec:edge}

Consider a globally hyperbolic manifold $\M$.\footnote{Or a asymptotically locally AdS spacetime with e.g. reflecting boundary conditions.} Let $\Sigma$ be a Cauchy slice at a moment of time symmetry, and denote by $a$ any open region of $\Sigma$, with $a' = \text{int}(\Sigma \setminus a)$ the complement of $a$ in $\Sigma$ minus the boundary of $a$. Also define the co-dimension 1 (co-dimension 2 in spacetime) {\it corner} of $a$ as $\partial a := \bar{a} \cap \bar{a}'$ where $\bar{a},\bar{a}'$ are the closures of $a,a'$. Intuitively, $\partial a$ is the boundary between $a,a'$. 

Let $\Ha_\Sigma$ be the Hilbert space of quantum gravity and matter defined on $\Sigma$. Because the topology of $\Sigma$ is fixed in this discussion, we expect our results to hold at best to all orders in perturbation theory. Of course, it is not even clear if the language of spacetime and Cauchy slices is a good approximation of the bulk physics in the non-perturbative regime, so an all orders statement may be the best we can hope for using these methods. The Hilbert space $\Ha_\Sigma$ may or may not be known explicitly. In AdS/CFT, $\Ha_\Sigma$ is isomorphic to a subspace of the Hilbert space of a CFT in one lower dimension \cite{Maldacena_1999}, but in flat or de Sitter spacetimes, less is known about the structure of $\Ha_\Sigma$.\footnote{See \cite{pasterski2021celestial,Pasterski_2021,strominger2018lectures,raclariu2021lectures} for work towards understanding this Hilbert space in flat space, and \cite{Strominger_2001,Balasubramanian:2001nb,witten2001quantum,chakraborty2023hilbert} for the case of de Sitter spacetimes.}
In the small Newton constant ($G_N$) limit, however, a great deal is known about $\Ha_\Sigma$: it contains perturbative gravitons and matter propagating on curved backgrounds. Take $\ket{\Psi} \in \Ha_\Sigma$ to be a state with a well-defined semi-classical description and a background geometry $g^0_{\mu\nu}$ in the $G_N \to 0$ limit. In this limit, we can define an (approximate) algebra of $\mathcal{A}_\Sigma$ of observables describing fluctuations around $\ket{\Psi}$ which act on $\Ha_\Sigma$ \cite{Almheiri_2015}. $\mathcal{A}_\Sigma$ includes both matter fields and perturbative gravitons $h_{\mu\nu}$ such that the total metric is $g_{\mu\nu} = g^0_{\mu\nu} + \sqrt{32\pi G_N} h_{\mu\nu}$, with the prefactor chosen to normalize the graviton kinetic term. Furthermore, $\Ha_\Sigma$ has a $\Diff(\Sigma)$ diffeomorphism gauge symmetry which corresponds to coordinate changes of the perturbative fields defined on $\Sigma$. In the language of AdS/CFT, the subspace spanned by $\mathcal{A}_\Sigma$ acting on $\ket{\Psi}$ defines a code subspace within the full non-perturbative Hilbert space of the bulk \cite{Almheiri_2015}. Our goal is to compute the entropy of the reduced density matrix $\rho_a \sim \tr_{a'} \ketbra{\Psi}$, associated with a bulk region $a$, in the weak coupling $G_N \to 0$ limit.
A precondition for this reduced state $\rho_a$ to be well-defined as a partial trace of the complementary subsystem is the existence of a bi-partition of the Hilbert space into $\Ha_\Sigma \sim \Ha_a \otimes \Ha_{a'}$. However, this factorization is subtle in quantum gravity for two distinct reasons.

First, there are UV divergences in the effective field theory (EFT) that we are considering, associated with the sharp cut $\partial a$ between $a,a'$ \cite{Haag:1992hx}.  We expect these divergences to be cured in the high energy completion defining $\Ha_\Sigma$ in something like string theory.  RG-flow from this complete theory leads to an EFT below a cutoff.  With this in mind, we will assume that appropriate regularization and renormalization makes the path integral that we will consider well-defined. Second, if there is a gauge symmetry in the theory, constraints can correlate $a$ and $a'$. This is even true for $U(1)$ lattice gauge theory \cite{Donnelly:2011hn}, and leads to a failure of the Hilbert space factorization between $a$ and $a'$. We review this lack of factorization in this section, following the discussions in \cite{Casini:2013rba,Donnelly:2011hn,Donnelly_2016,Balasubramanian:2025rcr,Balasubramanian:2026ymu,Balasubramanian:2026vlr}.

\subsection{Edge modes in lattice gauge theory}\label{sec:em_ym}

Consider a $U(1)$ lattice gauge theory, which manifestly has no UV divergences, and let $\mathcal{H}_\Sigma$ be the gauge invariant Hilbert space of the full lattice. As in the continuum, we separate the lattice into a subset of nodes $a$ and its complement $a'$, with $\partial a$ cutting the links between them.

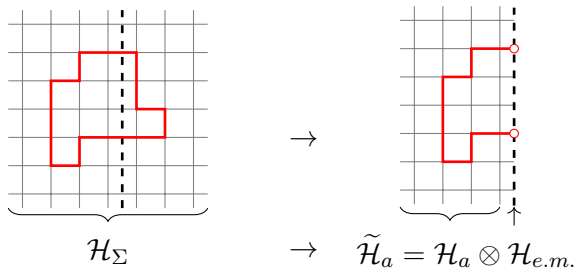
\begin{figure}
    \centering
    \begin{tikzpicture}[scale=0.75]
        \def\aa{1}
        \def\eps{1}
        \def\nn{7}
        \def\hh{3}
        \def\ss{0.5*0.75}
        
        \node at (-\hh,0) {
            \begin{tikzpicture}[scale=\ss]
                
                \draw[step=\aa,color=gray] (-0.5,-0.5) grid (\nn*\aa-0.5,\nn*\aa-0.5);
                \draw[line width=\eps,dashed]
                (\nn*\aa/2,-0.5)-- 
                (\nn*\aa/2,\nn*\aa-0.5);
                
                \draw [decorate,decoration={brace,amplitude=5pt,mirror}]
                (-0.5,-0.5) -- (\nn*\aa-0.5,-0.5) node[midway,yshift=-1.5em]{$\Ha_\Sigma$};
                
                \draw[red, line width=\eps] 
                (1,1) -- 
                (2,1) -- 
                (2,2) -- 
                (5,2) -- 
                (5,3) -- 
                (4,3) -- 
                (4,5) -- 
                (2,5) --
                (2,4) --
                (1,4) -- 
                cycle;
        \end{tikzpicture}};
        \node[anchor=west] at (0,0) {$\rightarrow$};
        \node[anchor=west] at (0,-1.954) {$\rightarrow$};
        \node at (\hh,0) {
            \begin{tikzpicture}[scale=\ss]
                
                \draw[step=\aa,color=gray] (-0.5,-0.5) grid (\nn*\aa/2 - \aa/2+0.5,\nn*\aa-0.5);
                \draw[line width=\eps,dashed]
                (\nn*\aa/2,-0.55)-- 
                (\nn*\aa/2,\nn*\aa-0.55);
                
                \draw[red, line width=\eps] 
                (3.5,5) -- 
                (2,5) --
                (2,4) --
                (1,4) -- 
                (1,1) -- 
                (2,1) -- 
                (2,2) -- 
                (3.5,2);
                
                \draw [decorate,decoration={brace,amplitude=5pt,mirror}]
                (-0.5,-0.6) -- (\nn*\aa/2-0.5,-0.6) node[midway,yshift=-1.5em]{\hspace{12.5pt}$\widetilde{\Ha}_a \subset \Ha_a \otimes \Ha_{\mathrm{edge}}$};
                \draw[->] (\nn*\aa/2,-1.25) -- (\nn*\aa/2,-0.6) ;
                
                \filldraw[red,fill=white] (3.5,2) circle (0.15); 
                \filldraw[red,fill=white] (3.5,5) circle (0.15); 
        \end{tikzpicture}};
    \end{tikzpicture}
    
    \caption{The full Hilbert space $\Ha_\Sigma$ is defined on the complete lattice. When restricting to a subset $a$ of the nodes on the lattice, described by a reduced Hilbert space $\Ha_a$, one must also include additional degrees of freedom to account for the endpoints of Wilson lines not present in $\Ha_\Sigma$. These new degrees of freedom are the edge modes in $\Ha_{\mathrm{edge}}$. The extended Hilbert space $\widetilde{\Ha}_a$ is the subspace of $\Ha_a \otimes \Ha_{\mathrm{edge}}$ such that the edge mode state transforms in the conjugate representation to the intersected links, so that right gauge transformations remain gauged.}
    \label{fig:lattice}
\end{figure}

Suppose a Wilson line straddles the corner $\partial a$, as in Fig.~\ref{fig:lattice}. Then tracing out $a'$ will leave behind an open Wilson line with endpoints on $\partial a$. From the perspective of $a$, the endpoints at the cut will carry charges transforming in some representation of the gauge group at $\partial a$. Since the original Hilbert space was gauge invariant, these open Wilson lines constitute new degrees of freedom  called ``edge modes''. The edge modes are a useful tool in computations of entanglement entropy in gauge theories \cite{Donnelly:2011hn}, or more generally, in systems whose operator algebras have a non-trivial center. Indeed, the total Hilbert space describing the degrees of freedom on $a$ is the ``extended Hilbert space'' $\widetilde{\Ha}_a \subset \Ha_a \otimes \Ha_{\mathrm{edge}}$, where $\Ha_a$ has support only inside $a$, and $\Ha_{\mathrm{edge}}$ is the edge mode Hilbert space supported on the corner. The extended Hilbert space, then, is the subspace of $\Ha_a \otimes \Ha_{\mathrm{edge}}$ such that the auxiliary edge mode degrees of freedom on the cut agree with representation of the bulk Wilson line which ends on the cut (which has support within $\Ha_a$). For instance, $\widetilde{\Ha}_a$ does not include non-trivial edge modes with no Wilson lines to support them.

To explain how to recover the original Hilbert space, we must first understand the extended Hilbert space. The total gauge symmetry of the lattice $\Sigma$ is $U(1)_{\Sigma}$, i.e., $U(1)$ transformations at every link on the lattice. This gauge group splits as \begin{align}
    U(1)_{\Sigma} = U(1)_a \oplus U(1)_{\partial a} \oplus U(1)_{a'}\,,
\end{align} 
where $U(1)_{\partial a}$ are gauge transformations transforming links along the cut, while $U(1)_a$ consists of gauge transformations of links with both endpoints in $a$ (and similarly for $U(1)_{a'}$).\footnote{More precisely, gauge transformations are defined at the vertices of the lattice. A gauge transformation at a vertex $v$ acts on every edge which ends at that vertex. When we split the gauge group in this way, we really mean $U(1)_a$ consists of all gauge transformations which act on the vertices of $a$, and by $\partial a$ we have placed a vertex bisecting every link $\partial a$, and we mean the gauge transformations on these vertices. Because of the gauge invariance of $\Ha_\Sigma$, there is no physical difference between the lattice with vertices on $\partial a$ and that without. But these details will not be important for our analysis in this paper.
} 
The extended Hilbert space $\widetilde{\Ha}_a $ has a gauge symmetry $ U(1)_a$ that it inherits from $\Ha_a$, and a \emph{global} symmetry $U(1)_{\partial a}$ acting on $\Ha_{\mathrm{edge}}$. The latter is a global symmetry because the open Wilson lines transform in a non-trivial representation of $U(1)_{\partial a}$. In fact, a more careful consideration shows that $\Ha_{\mathrm{edge}} \cong L^2(U(1)_{\partial a})$.\footnote{To see this, first note that the edge mode at each link is simply a choice of phase \cite{Donnelly_2016}. Thus, if there are $n$ links on the cut $\partial a$, $\Ha_{\mathrm{edge}} \cong L^2(U(1))^{\otimes n} \cong L^2(U(1)^{\otimes n}) \equiv L^2(U(1)_{\partial a})$ because $L^2(A) \otimes L^2(B) \cong L^2(A \otimes B)$.
}
This makes the global symmetry more transparent: it is simply the left multiplications of $U(1)_{\partial a}$ on $L^2(U(1)_{\partial a})$. 
We can reason similarly about $\widetilde{\Ha}_{a'}$.

Because the extended Hilbert space has a global symmetry $U(1)_{\partial a}$, it splits into superselection sectors of charge eigenstates. The representations of $U(1)_{\partial a}$ are given by $\vec{Q}$, a choice of $U(1)$ charge for every link on the cut, and so $\widetilde{\Ha}_a = \oplus_{\vec{Q}} \widetilde{\Ha}_{\vec{Q},a}$. Thus, the most general density matrix in $\widetilde{\Ha}_a$ which descends from a gauge invariant state on $\Ha_\Sigma$ takes the form $\rho_a = \sum_{\vec{Q}} p_{\vec{Q}} \rho_{\vec{Q}}$, where $\sum_{\vec{Q}} p_{\vec{Q}} = 1$ and each charge-sector density matrix has unit trace. $\rho_{\vec{Q}}$ is any state compatible with the boundary charge $\vec{Q}$. For example, $\rho_{\vec{Q}=0}$ is a possibly mixed state over Wilson loops which do not intersect the cut. We will return to the implications of this splitting below.

Let us now consider how to recover the original Hilbert space $\Ha_\Sigma$. First, note that $\Ha_a \otimes \Ha_{a'} \subset \Ha_\Sigma$ because $\Ha_a \otimes \Ha_{a'}$ does not contain any Wilson lines straddling the cut ${\partial a}$. This was why we had to introduce the edge modes in the first place. On the other hand, $\Ha_\Sigma \subset \widetilde{\Ha}_a \otimes \widetilde{\Ha}_{a'}$ because $\widetilde{\Ha}_a \otimes \widetilde{\Ha}_{a'}$ contains states with open Wilson lines on the cut, but $\Ha_\Sigma$ only contains closed Wilson loops. We can recover $\Ha_\Sigma$ if we restrict to states with a trivial charge under $U(1)_{\partial a}$, that is, \begin{align}
    \Ha_\Sigma = \Pi_0 [\widetilde{\Ha}_a \otimes \widetilde{\Ha}_{a'}] \,,
\end{align}
where $\Pi_0$ projects states onto the subspace satisfying $\vec{Q}_a + \vec{Q}_{a'}=0$. We now see why the total Hilbert space is gauge invariant under the full $U(1)_{\Sigma}$ even if it is built from states with global charges in $\widetilde{\Ha}_a,\widetilde{\Ha}_{a'}$. While $\widetilde{\Ha}_a \otimes \widetilde{\Ha}_{a'}$ has a global symmetry group $U(1)_{\partial a} \oplus U(1)_{\partial a}$, restricting to the zero charge subspace identifies the overlapping links on the cut with each other and gauges the remaining global symmetry. Alternatively, $\Ha_\Sigma$ is not the same as $\Ha_a \otimes \Ha_{a'}$ because the latter is missing the twisted sectors required for a consistent gauging of a global symmetry \cite{Vafa:1989ih,Dijkgraaf:1989hb}.

The abelian criterion $\vec{Q}_a + \vec{Q}_{a'}=0$ indicating that we have gauged $U(1)_{\partial a}$ has a non-abelian generalization. Let $G$ be a compact Lie group for a non-abelian lattice Yang-Mills theory. Then the full gauge group splits in the same way as $U(1)$: $G_{a{a'}} = G_a \oplus G_{\partial a} \oplus G_{a'}$.
Even in the non-abelian case $\Ha_{\mathrm{edge}} \cong L^2(G_{\partial a})$.\footnote{The edge mode corresponds to a choice of trivialization of the gauge function at the boundary \cite{Donnelly_2016}. Thus, the proof works exactly the same as in the Abelian case.
} Because of this, we can utilize the Peter-Weyl theorem \cite{PeterWeyl}, which demonstrates that $L^2(G_{\partial a})$ decomposes as
\begin{align}
    L^2(G_{\partial a}) \cong \bigoplus_R \, d_R \cdot[V_{R} \otimes V_{\overline{R}}] \, , \label{eqn:peterweyl}
\end{align}
where $V_{R}$ is the vector space of the irrep $R$ with dimension $d_R$. The notation $d_R \cdot [ \cdots ]$ denotes a relative rescaling of the inner product on each vector space. In other words, a unit normalized vector $\ket{R,ij}$ in $V_R \otimes V_{\overline{R}}$ should be assigned a norm of $d_R$ if we want to view  $V_R \otimes V_{\overline{R}}$ as a subspace of $L^2(G_{\partial a})$. Of course, this rescaling is not necessary if we only care about the equality \eqref{eqn:peterweyl} at the level of vector spaces. But if we want \eqref{eqn:peterweyl} to hold at the level of \emph{Hilbert} spaces, the rescaling is required. 

More generally, for non-compact semisimple Lie groups,\footnote{Even more generally, any locally compact, separable, unimodular group whose irreducible representations are all type I algebras.} we can use the Plancherel decomposition
\begin{align}
    L^2(G_{\partial a}) = \int_{\widehat{G}} d\mu(R) V_R \otimes V_{\overline{R}} \,.
\end{align}
The Plancherel decomposition is similar to the Peter-Weyl theorem for compact groups, but can account for a continuum of possible representations (for example, the continuum of momenta for $G=\R$). Analogously to the rescaling $d_R\cdot[\cdots]$ (which is actually a special case of the Plancherel decomposition), the Plancherel measure is the unique measure which makes the Fourier transform between the group basis and the representation basis a unitary map. See \cite{Balasubramanian:2025rcr} for a brief review of the representation theory of non-compact Lie groups. For the rest of this paper, we will use the Plancherel integral notation, with the understanding that $d\mu(R)$ is delta function localized (with coefficient $d_R$) if $G$ is a compact group.

The extended Hilbert space decomposes as
\begin{align}
    \widetilde{\Ha}_a = \int_{\widehat{G}_{\partial a}} d\mu(R) \, \Ha_a(R) \otimes \Ha_{\mathrm{edge}}(\overline{R}) \,. \label{eq:exthilb}
\end{align}
The fact that we only pair bulk fields with boundary conditions transforming in the $R$ representation with edge modes transforming in the conjugate $\overline{R}$ representation ensures that right multiplication by the gauge group remains a gauge symmetry in the extended Hilbert space. Bulk gauge transformations $G_a$ are also a gauge symmetry in the extended Hilbert space. Left multiplications by $G_{\partial a}$, however, are a global symmetry, which acts on the endpoints of open Wilson lines ending on the cut. These global symmetries only act on $\Ha_{\mathrm{edge}}(R)$.

As in the abelian case, we can recover the global Hilbert space $\Ha_\Sigma$ by taking a subspace of the extended Hilbert spaces of $a$ and $a'$ as follows. Let $g \in G_{\partial a}$ be a non-abelian gauge transformation on the cut. Given a state $\ket{\psi_a} \otimes \ket{\psi_{a'}} \in \widetilde{\Ha}_a \otimes \widetilde{\Ha}_{a'}$, we recover a gauge-invariant state on the full Cauchy slice by 
projecting onto the equivalence classes under the diagonal action of $G_{\partial a}$. That is, we identify $\ket{\psi_a} \otimes \ket{\psi_{a'}} \sim g\ket{\psi_a} \otimes g\ket{\psi_{a'}}$ as being the same state. This projection is referred to as the ``entangling product'' \cite{Donnelly_2016}, or the Hilbert space of co-invariants \cite{Chandrasekaran:2022cip,Balasubramanian:2025rcr}, and is sometimes denoted as
\begin{align}
    \Ha_\Sigma = \widetilde{\Ha}_a \otimes_{G_{\partial a}} \widetilde{\Ha}_{a'} \,.\label{eqn:entprod}
\end{align}
The entangling product ensures continuity of the bulk fields across the cut after gluing the two Hilbert spaces together.

Let us now understand the structure of a reduced state $\rho_a$, thought of as an operator acting on the extended Hilbert space $\widetilde{\Ha}_a$. For simplicity, momentarily take $G$ to be a compact gauge group, and let $R$ be a unitary irreducible representation (irrep) of $G_{\partial a}$. Since $G_{\partial a}$ is a global symmetry for the extended Hilbert space (equivalently, from the direct integral decomposition of \eqref{eq:exthilb}), the most general state in $\widetilde{\Ha}_a$ in the non-abelian case would seem to be given by $\rho_a = \sum_R p_{R} \, \rho_R \otimes \sigma_R$, where $\rho_R$ and $\sigma_R$ are density matrices on $\Ha_a$ and $\Ha_{\mathrm{edge}}$, respectively. Here $\rho_R$ is a possibly mixed state over all bulk states compatible with a representation $R$ on the cut, similar to the $U(1)$ case above. 

Naively, $\sigma_R$ seems to be unconstrained. However, let $\rho_\Sigma$ be a state in the gauge invariant Hilbert space (\ref{eqn:entprod}). Because $\Ha_\Sigma \subset \widetilde{\Ha}_a \otimes \widetilde{\Ha}_{a'}$, we can embed $\rho_\Sigma$ into $\widetilde{\Ha}_a \otimes \widetilde{\Ha}_{a'}$ with an isometry $V$ and define 
\begin{align}
    \rho_a := \tr_{\widetilde{\Ha}_{a'}} [V\rho_\Sigma V^\dagger] \,,
\end{align}
where the trace is taken in the larger $\widetilde{\Ha}_a \otimes \widetilde{\Ha}_{a'}$ because this is the Hilbert space with a well-defined tensor product factorization. We will return to the definition of this trace below. Because $G_{\partial a}$ is a gauge symmetry of $\Ha_\Sigma$, $\rho_a$ must commute with the generators of $G_{\partial a}$. 
By Schur's lemma, this implies that $\sigma_R$ is proportional to the identity for each $R$. Therefore, when $G$ is a compact group, the most general density matrix which arises from a globally gauge invariant pure state takes the form \cite{Donnelly:2011hn}
\begin{align}
    \rho_a = \sum_{R\in\widehat{G}_{\partial a}}  p_{R} \, \rho_{R} \otimes \frac{1}{d_R} \Id_{R} \, , 
\end{align}
where $\Id_{R}$ is the identity in representation $R$.
If $\rho_\Sigma$ was properly normalized, then so is $\rho_a$, which corresponds to the conditions
\begin{align}
    \sum_{R} p_R = 1\,, && \tr_R[\rho_R] = 1\,,
\end{align}
where $\tr_R$ is the usual trace on $\Ha_a(R)$.
When $G$ is a non-compact group (the case most immediately relevant for gravity), $d_R$ is infinite, and $\widehat{G}_{\partial a}$ has a continuum of representations. In that case, we instead have the decomposition of $\rho_a$ as \cite{Balasubramanian:2026ymu,Balasubramanian:2026vlr} 
\begin{align}
    \rho_a = \int_{\widehat{G}_{\partial a}} d\mu(R) q_{R} \, \rho_{R} \otimes \Pi_R \,, \label{eq:rho_a}
\end{align}
where $\Pi_R$ is the Plancherel measure normalized projector
\begin{align}
    \Pi_R \Pi_{R'} = \delta(R,R') \, \Pi_R\,.
\end{align}
Here, $\delta(R,R')$ is the delta function adapted to the Plancherel measure $d\mu(R)$. Indeed, because $d\mu(R) \propto d_R$ when $G$ is compact, $\Pi_R = \Id_R / d_R$ in that case. So $\Pi_R$ can be thought of as the generalization of the maximally mixed state when $G$ is non-compact. Furthermore, $q_R$ is a probability distribution which is normalized with respect to the Plancherel measure. So in the compact group case, we would have that $p_R = d_R \, q_R$.

In order for $\rho_a$ to be a legitimate density matrix, it must be normalized to have unit trace. If we used the naive trace for arbitrary bounded operators on $\widetilde{\Ha}_a$, i.e. the ``usual definition'' of trace, then $\Tr[\rho_a] = \infty$ because $\Pi_R$ is delta function normalized. However, $\rho_a$ is \emph{not} an arbitrary operator, because it always block diagonalizes as above. For operators of this form, there is an alternative definition of trace for which $\rho_a$ is unit normalized. Because these complications only arise for theories with non-compact gauge groups, of which gravity is a key example, we will defer the details of this trace to Sec.~\ref{sec:entropy}.

\subsection{Edge modes in gravity} \label{sec:em_grav}

Gravity is a gauge theory for the diffeomorphism group, and thus also has edge modes \cite{Donnelly_2016}, but there are some key differences from standard gauge theory that are worth emphasizing. As described above, associated with a Cauchy slice $\Sigma$ is a Hilbert space $\Ha_\Sigma$, with a gauge group $\Diff(\Sigma)$, spacetime diffeomorphisms with support on $\Sigma$ (including timelike translations of the slice). Given subregions $a,a'$, induced by the corner $\partial a$, we would like to bipartition the Hilbert space into $\Ha_\Sigma$ into two tensor factors, one for $a$ and one for $a'$. There are again two issues with this bipartition: UV divergences and constraints from the $\Diff(\Sigma)$ gauge symmetry. As discussed above, we assume that the UV divergences have been regulated and renormalized in some way, and will not discuss this point in detail in this paper.

Because $\mathrm{Diff}(\Sigma)$ is not a locally compact group, we can not immediately apply the technology of Yang-Mills theory to gravity.\footnote{Local compactness is necessary for the existence of the Haar measure, and by Fourier duality, the existence of the Plancherel measure. However, as explained below, we can still formally define the Plancherel measure for the full diffeomorphism group as the Fourier dual of a path integral measure, rather than a genuine integration measure. But because a path integral measure is often defined as a limiting sequence of integration measures (i.e., in the limit the lattice spacing vanishes), we should be able to use this formalism for any finite cutoff, and take the limit at the end.}
However, recall that we are treating gravity as an EFT above some cutoff $\Lambda$. Included in this regulator is a deformation of $\mathrm{Diff}(\Sigma)$. For example, if we integrate out the large momentum modes above some cutoff $\Lambda \sim M_P$, then diffeomorphisms which only move points within length scales $\Lambda^{-1} \sim \ell_P$ must be quotiented out of the gauge group, for they have no degrees of freedom to act upon. Thinking of $\mathrm{Diff}(\Sigma)$ as local translations of $\Sigma$ within the spacetime $\mathcal{M}$, this effectively acts as a UV regulator for the gauge group itself. So if we approximate $\mathrm{Diff}(\Sigma)$ by some locally compact subgroup, or some other deformation of $\mathrm{Diff}(\Sigma)$ which renders it locally compact, this approximation should only break down below the Planck scale. Therefore, we will formally treat $\Diff(\Sigma)$ as a locally compact group, and we expect our analysis to be a good approximation for observables which are not sensitive to Planck scale physics. 

This also shows that we should expect the deformation itself to depend on $G_N$, because the Planck length scales as $\ell_P \sim G_N^{1/(d-2)}$. Therefore, the representation data such as the Plancherel measure should also carry a $G_N$ dependence. 
An example of this phenomenon is in 3D gravity in AdS spacetimes.\footnote{We will work in units where the AdS radius has been set to unity.} In the first order formalism, the $\SL(2,\R)$ gauge symmetry of each chiral half (whose gauge transformations are on-shell equivalent to diffeomorphisms \cite{Witten:1988hc})
is modified to the modular double of the quantum group $\mathcal{U}_q(\sl(2,\R))$, where $q = \exp(4\pi i G_N )$ for small $G_N$.\footnote{This follows because $q = \exp(i \pi b^2)$ \cite{Ponsot:1999uf}, where $c=1+6(b+b^{-1})^2$ is the Brown-Henneaux central charge $c = \frac{3 \ell}{2 G_N}$ \cite{Brown:1986nw}.} So the deformation parameter of the gauge symmetry indeed depends on $G_N$ in this case, and thus so will the representation theory of the deformed diffeomorphism group. We will leave this deformation implicit in what follows, and let the Plancherel measure $d\mu(R)$ be this deformed measure.

To deal with the gravitational constraints, we first decompose the gauge group as 
\begin{equation}
    \Diff(\Sigma) = \Diff(a) \oplus \text{ECS} \oplus \Diff(a')\,.
\end{equation}
Here, $\Diff(a)$ and $\Diff(a')$ are diffeomorphisms with support within $a,a'$ respectively and vanishing on the corner.
ECS is the equivalence class of the remaining diffeomorphisms, called the ``extended corner symmetry'' group\footnote{See \cite{Ciambelli:2022cfr} for more information about the importance of this group in gravity.} \cite{Freidel:2021cjp}. ECS has a Lie algebra given by
\begin{equation}
    \ecs = \diff(\partial a) \ltimes (\sl(2,\R) \ltimes \R^2)_{\partial a}    
\end{equation}
independently of the size and topology of $a$. The  $\diff(\partial a)$ algebra describes coordinate reparameterizations of the corner. Now let $p \in \partial a$ be a point on the corner and $x^\pm$ be coordinates on the two dimensional plane normal to $\partial a$ that intersects $p$, such that $p$ is defined by $x^\pm = 0$. One can think of this as the plane spanned by two null rays leaving $p$. The $\sl(2,\R)$ algebra describes transformations of this plane which keep $p$ fixed and have unit determinant. This includes boosts around $p$, but other transformations are possible as well. The $\R^2$ part generates translations of the point $p$ itself:\footnote{This is the analog of supertranslations in the BMS group at infinity in flat space.} together, $\sl(2,\R) \ltimes \R^2$ forms a kind of Euclidean group of the normal plane. The notation $( \cdots )_{\partial a}$ means that we are actually considering a copy of this Euclidean group at each point $p$.\footnote{More precisely, it is the symmetry group of the normal bundle over $\partial a$.} Remarkably, this charge algebra has no central extensions, which simplifies its analysis \cite{Ciambelli:2021nmv}.

If gravity were just like Yang-Mills theory, only $\diff(\partial a)$  would appear. The presence of the extra terms $ (\sl(2,\R) \ltimes \R^2)_{\partial a}$ has no analog in Yang-Mills, and corresponds to diffeomorphisms which affect the time and radial coordinates describing the slice $\Sigma$. In the lattice model, $\diff(\partial a)$ corresponds to gauge transformations on each link, but $ (\sl(2,\R) \ltimes \R^2)_{\partial a}$ captures changes of the lattice itself. This is a manifestation of the famous ``problem of time'' in quantum gravity, because $\diff(\partial a)$ all keeps the corner fixed, while the extra $ (\sl(2,\R) \ltimes \R^2)_{\partial a}$ factors include time translations of the corner within the global spacetime.
 
\begin{figure}
    \centering
    \includegraphics[]{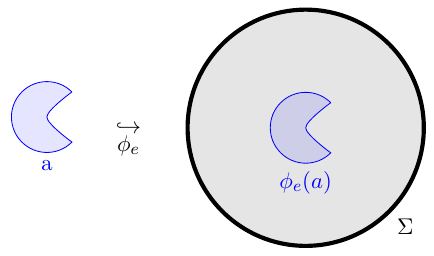}
    \caption{$a$ is a manifold with boundary, independent of $\Sigma$. By specifying an embedding $\phi$, we can think of $a$ as a subset of $\Sigma$ in a gauge invariant way. Throughout this paper, we refer to $\phi_e(a)$ as $A$.}
    \label{fig:embed}
\end{figure}

Next, we should define more clearly what we mean by a subregion $a$ in a diffeomorphism invariant system. Let $\partial a$ be the corner separating $a$ and $a'$. Since diffeomorphisms are a gauge symmetry of gravity, simply giving the coordinate values for $\partial a$ is not enough to define the corner, as this is not a gauge invariant quantity without also fixing the coordinate system of the spacetime, i.e., choosing a gauge. However, we can think of $a \subset \Sigma$ as an open, embedded submanifold of $\Sigma$, and this perspective does not depend on a choice of coordinates for $\Sigma$. Recall that an embedded submanifold is defined by a subset of $\Sigma$ (here, the subregion $a$), as well as a choice of embedding map $\phi: a \into \Sigma$. In other words, we are viewing $a$ as an open manifold in its own right, and the embedding map $\phi$ provides a definition for how to map points in $a$ into $\Sigma$. Explicitly keeping track of $\phi$ as a dynamical field, essentially a Stueckelberg field for diffeomorphisms,\footnote{See \cite{geng2023stueck} for an example of how Stueckelberg fields for gravity have appeared elsewhere in the literature.} as well as the fields on the fixed model subregion $a$ is therefore sufficient for specifying a subregion $a \into \Sigma$, as shown in Fig.~\ref{fig:embed}. 
Such an embedding also restricts to an embedding of the corner $\phi(\partial a) \into \Sigma$ into the Cauchy slice. In the same way, there is a complementary embedding $\phi'$ of $a'$ into $\Sigma$. Because $\partial a = \partial a'$ up to a choice of orientation, the embedding maps must be compatible: $\phi(\partial a) = \phi'(\partial a')$. This restriction is what ensures the gluing between the two complementary subregions remains compatible, even after a diffeomorphism.

We will abuse notation and refer to this abstract subregion as $a$ from now on, rather than the embedding of $a$ within $\Sigma$. We then view the original region in $\Sigma$ as a particular embedding $\phi_e: a \into \Sigma$, and define $A_e := \phi_e(a)$. 
The physical quantum fields $\psi|_\Sigma$ which are restricted to the slice $\Sigma$ are defined on this subregion: in other words, we label the fields by their values $\psi(\phi(x))$ (which is a diffeomorphism invariant specification of the field value), rather than $\psi(x)$.
At the same time, the complementary subregion is defined by $A_e' = \phi'_e(a')$. So specifying the embedding map $\phi_e$ is a diffeomorphism invariant way to define the cut between $a$ and $a'$. This is because a diffeomorphism of $\Sigma$ will also transform the map $\phi$ to keep the split between $a$ and $a'$ invariant.  In fact, this $\phi$ is precisely the edge mode of gravity \cite{Ciambelli:2021vnn,Ciambelli:2021nmv}. To see this, let $\phi_e:a \into \Sigma$ be an embedding. Then, let $g \in \text{ECS}$ be a diffeomorphism. Then $\phi_g := g \cdot \phi_e$ is also an embedding, and distinct from $\phi_e$ unless $g = e$. It is easy to see from this definition that the set of $\phi_g$ form a representation of ECS, as do the edge modes, so it is plausible that this is a correct way to view them. A more careful analysis reveals this to indeed be the case \cite{Ciambelli:2021vnn,Ciambelli:2021nmv}. 

We emphasize that while $\Sigma = A_g \cup_{\partial a} A_{g^{-1}}'$ for any $g \in \mathrm{ECS}$, $A_e \neq A_g$ in general. For example, consider Minkowski space with a fixed choice of Cartesian coordinates $(t,x,\vec{y}_\perp)$, with $\vec{y}_\perp$ a collection of 
$d-2$ transverse coordinates, which is a choice of gauge for the diffeomorphism symmetry.\footnote{For this particular state, this only fixes the metric up to an isometry. However, isometries are large diffeomorphisms, and so are a global symmetry on a fixed background. Because we are not focusing on topology change in this paper, this choice of gauge is sufficient for our purposes.}
Let $\Sigma$ be the Cauchy slice specified by $t=0$ in this coordinate system. Then one choice of subregion $a$ is the half space $x < 0$, and the subregion $a'$ would be the complementary half space $x > 0$. However, these subspaces can be viewed as manifolds in their own right, so we can act diffeomorphisms on each subregion independently. For example, let $s(y_\perp)$ be a compactly supported bump function. Then we can transform $a$ by the diffeomorphism $f: x \to x + s(y_\perp) e^{- \epsilon x^2}$ on the $x$ coordinate of the $a$ region, for any $0 < \epsilon < 1$. 
This diffeomorphism is small with respect to the asymptotic boundary, and acts as a translation $x \to x+s(y_\perp)$ near the corner, so it is large with respect to the corner. The corner $f(\partial a)$ of $f(a)$ shifts from the coordinate value $x=0$ to $x=s(y_\perp)$. All fields, including the metric, transform under this diffeomorphism.
However, the embedding map $\phi$ also transforms under this diffeomorphism as $\phi_f = \phi_e \circ f^{-1}$. So as submanifolds of $\Sigma$, $\phi_e(a) = \phi_f(f(a))$, and so the pair $(a,\phi)$ is a diffeomorphism invariant specification of the Rindler wedge. However, the submanifold $\phi_f(a)$ is \emph{not} the same as $\phi_e(a)$. For example, tracking the definitions, $\phi_f(a) \cap \phi_e'(a')$ is the subregion of $\Sigma$ which is bounded by $x=0$ and $x=s(y_\perp)$. This shows that the embedding map can genuinely move the subregion within the spacetime. 
However, if we transform $\phi$ and $\phi'$ in complementary ways, then $\Sigma$ is preserved: for any $f \in \mathrm{ECS}$, $\Sigma = \phi_f(a) \cup_{\partial a} \phi_{f^{-1}}'(a')$, because the corner is translated in compatible ways on both sides of the cut. 

There are two ways to view the subregion $\phi_f(a)$. First, as explained above, we can view $\phi_f$ as a genuinely new embedding of the same subregion $a$. From the perspective of the subregion $a$, this is a ``passive'' view of the diffeomorphism $f$: the coordinate values defining the subregion change, but the underlying manifold is itself unchanged. Alternatively, because $\phi_f(a) = \phi(f^{-1}(a))$, we can equivalently view $\phi_f(a)$ as fixing the coordinate value of the cut, but transforming the fields within the subregion $a$ according to the diffeomorphism $f$, including the metric. This is an ``active'' view of the diffeomorphism $f$. Because $\phi_f(a) = \phi(f^{-1}(a))$, the two views are completely equivalent. 
Together, analogous to the gauge theory example above, we can recover the global state in gravity from the pairs $(a,\phi)$ and $(a',\phi')$, subject to the gluing constraint $\phi(\partial a) = \phi'(\partial a')$. Quotienting out by the bulk diffeomorphisms $\mathrm{Diff}(a) \sqcup \mathrm{Diff}(a')$, only the embedding of the corner is a physical degree of freedom.
We should then quotient the glued state by the diagonal action of $\mathrm{ECS}$ on the embeddings, because this preserves the state, and is therefore redundant. This is exactly the same structure as the entangling product of the gauge theory example above.

That said, the fact that diffeomorphisms can move the subregion within the spacetime has no analog within Yang-Mills theory, and is highly constraining on the quantum state of gravity. For example, in the Rindler wedge example, the subregions $a$ and $a'$ are only complementary subregions when we gauge transform both sides of the cut. By a diffeomorphism, we can shift the definition of $a$ to include regions of the spacetime originally included in the complementary subregion $a'$. So it seems like the extended Hilbert space $\widetilde{\Ha}_a$ must know more about the global spacetime than just the subregion $a$ we started with. 

\subsection{The reduced state} \label{sec:reducedstate}

In analogy with Yang-Mills theory, we propose that the extended Hilbert space in gravity is explicitly given by
\begin{align}
    \widetilde{\Ha}_a = \int_{\widehat{ECS}} d\mu(R) \Ha_a(R) \otimes \Ha_{\mathrm{edge}}(\overline{R}) \,. \label{eq:exthilb2}
\end{align}
Here, $\Ha_a(R)$ is the bulk Hilbert space associated with field configurations with boundary conditions that transform in the $R$ representation of the corner symmetry group. $\Ha_{\mathrm{edge}}(\overline{R})$, then, is the edge mode Hilbert space of the embedding degrees of freedom which transform in the conjugate representation. This pairing is forced by the constraint that right multiplications by diffeomorphisms remain a gauge symmetry of the extended Hilbert space. Furthermore, we have assumed that $\mathrm{ECS}$ has been UV regulated to be locally compact, and leave the choice of regulator implicit.

Given the decomposition \eqref{eq:exthilb2} of the regulated Hilbert space, as in Yang-Mills theory, any density matrix $\rho_a$ which descends from a global state in $\Ha_\Sigma$ can be written in the form \cite{Balasubramanian:2026ymu}
\begin{equation}
    \rho_a = \int d\mu(R) \,q_R\, \rho_R \otimes \Pi_R \,. \label{eq:reducedstate}
\end{equation}
Here, $d\mu(R)$ is the Plancherel measure for the (regulated version of) ECS, $\rho_R$ is a reduced state for the bulk fields in the $\Ha_a(R)$ factor of the extended Hilbert space, and $q_R$ is a probability distribution (normalized with respect to the Plancherel measure) over the choice of representations of $\mathrm{ECS}$. Furthermore, $\Pi_R$ is a projector onto the $R$ representation of $\mathrm{ECS}$, which satisfies
\begin{align}
    \Pi_R \Pi_{R'} = \delta(R,R') \Pi_R \label{eq:Piproj}
\end{align}
where $\delta(R,R')$ is the Plancherel-normalized delta function on the representations $R$.

Each $\rho_{R}$ can be interpreted as a kind of ``fixed area state'' \cite{Dong:2018seb} of the bulk subregion. Indeed, if we focused on a state where the expansion $\theta_\pm$ vanishes at all points along $\partial a$, then $\partial a$ is an extremal surface within the spacetime. In this case, one could use this condition to fix the corner $\partial a$ in a gauge invariant way, and therefore imagine neglecting the diffeomorphisms which move the surface. This is exactly what the authors of \cite{Dong:2018seb} do: they show that the resulting extended phase space of General Relativity consists precisely of an induced metric on the surface and a choice of conformal structure on the null planes normal to the corner. These are the edge modes of the $\mathrm{Diff}(\partial a)$ and $\SL(2,\R)$ components of $\mathrm{ECS}$, respectively. The area of the corner labels the eigenvalues of the Casimir of this representation \cite{Donnelly:2020xgu}, so this structure is consistent with their analysis. The role of the extended Hilbert space is to instead quantize the \emph{maximal} symmetry algebra on the corner, rather than the subset which fixes the corner. We use edge modes in our analysis of subregions, but the analogous operator algebra approach of \cite{Dong:2018seb} (upgraded to include the normal translations of the corner) should be equivalent. 

Therefore, let $V : \Ha_\Sigma \to \widetilde{\Ha}_a \otimes \widetilde{\Ha}_{a'}$ be an embedding of the global Hilbert space into the extended Hilbert spaces. Then, let $\ket{\Psi} \in \Ha_\Sigma$ be a state in the global Cauchy slice. We can then define the reduced state
\begin{align}
    \rho_a = \Tr_{a'}[V \ketbra{\Psi} V^\dagger]\,.
\end{align}
Our goal is now to compute the entropy
\begin{align}
    S(\rho_a) = -\Tr_{a}[ \rho_a \ln(\rho_a) ]\,.\label{eqn:entropy_rholnrho}
\end{align}
By $\Tr_{a}$, we mean the renormalized trace such that for an operator $\mathcal{O}$ which acts on the extended Hilbert space as
\begin{align}
    \mathcal{O} = \int d\mu(R) \mathcal{O}_R \otimes \Pi_R \,,
\end{align}
($\mathcal{O}_R$ is an operator that acts on $\Ha_a(R)$) the renormalized trace is defined as
\begin{align}
    \Tr_{a}[\mathcal{O}] := \int d\mu(R) \tr_R[\mathcal{O}_R]\,, \label{eq:physicaltrace}
\end{align}
where $\tr_R$ is the unique trace on each factor $\Ha_a(R)$. This is the trace over representations which arises if we used the Haar measure over $\mathrm{ECS}$ in the group basis to define the inner product for the extended Hilbert space. Using the terminology of \cite{Balasubramanian:2026ymu,Balasubramanian:2026vlr}, we call this the \emph{physical trace}.\footnote{In the notation of \cite{Akers:2023fqr}, this is analogous to the generalized trace.} If we chose a different measure over group elements to define the inner product, then the physical trace would be multiplied by a representation-dependent factor (the Fourier transform of the function multiplying the Haar measure). So this choice of trace is not unique, but to the extent the Haar measure over the embedding fields is preferred (as is the standard choice for most other quantum fields), the physical trace is preferred. Different choices of trace can be related by a choice of central operator
\begin{align}
    \mathcal{C} = \int d\mu(R) \,C(R) \cdot \Id_{\Ha_a(R)} \otimes \Pi_R \,. \label{eq:centralop}
\end{align}
Here, $C(R) \in \C$ are a set of coefficients defining the operator. All possible definitions of trace on the extended Hilbert space can be related as
\begin{align}
    \Tr^{\mathcal{C}}_{a}[\mathcal{O}] =  \Tr_{a}[\mathcal{C}^{-1} \mathcal{O}] \,. \label{eq:traceparambycenter}
\end{align}
The trace must be a positive real number for any positive operator such as $\mathcal{O}^\dagger \mathcal{O}$, so the coefficients $C(R)$ of the central operator $\mathcal{C}$ must be positive real numbers in \eqref{eq:traceparambycenter}.
In this notation, the physical trace corresponds to the choice $\mathcal{C} = \Id_{\widetilde{\Ha}_a}$. But the more invariant specification of the physical trace is that we integrate each trace $\tr_{R}$ using the Plancherel measure over each representation. 

\paragraph{The microcanonical trace.}

Before we compute the entropy of $\rho_a$, we must discuss the role of operators with support on the corner. By the expansion \eqref{eq:exthilb2} of the extended Hilbert space, an operator which only has support on the corner will commute with any operator in the bulk of the subregion $a$, i.e., it will act as the identity on each $\Ha_a(R)$. On the other hand, any gauge invariant operator must have an representation on the unfactorized Hilbert space $\Ha_\Sigma$, which only contains the $\Ha_a(R)$ factors. The central operators $\mathcal{C}$ discussed above are precisely the operators which satisfy both conditions, so we can think of \eqref{eq:centralop} as parameterizing the physical operators which have support on the corner.

Below, it will be useful to isolate the contribution of the central operators to the entropy of $\rho_a$. Indeed, for any state $\rho_a$ of the form \eqref{eq:reducedstate}, and any operator of the form \eqref{eq:centralop} with bounded coefficients $C(R)$, the expectation value 
\begin{align}
    \langle \mathcal{C} \rangle_\rho = \Tr_a[\rho_a \mathcal{C}] = \int d\mu(R) q_R C(R)
\end{align}
is finite. Because $\mathcal{C}$ acts solely on the corner, we could also measure $\mathcal{C}$ from the complementary subregion, and obtain the same finite expectation value.
Intuitively, there should therefore be a contribution to the entropy of $\rho_a$ arising from the correlated measurement outcomes of central operators $\mathcal{C}$ between the subregion $a$ and its complement. To make this precise, let $\mathcal{A}_a$ be the algebra of operators of the form $\int d\mu(R) \mathcal{O}(R) \otimes \Pi_R$, and let $\mathcal{C}_a$ denote the center of $\mathcal{A}_a$, i.e., the operators of the form \eqref{eq:centralop}. We can define a map $\mathcal{F}_\lambda$ by
\begin{align}
    \mathcal{F}_\lambda(\rho_a) = \int d\mu(R) \lambda(R) q_R \,\cdot \Id_{\Ha_a(R)} \otimes \Pi_R \,,
\end{align}
where $\lambda(R)$ is a real, strictly positive function of the representations $R$. This map essentially isolates the distribution $q_R$ which controls the expectation value of central operators. Such a map is an example of an ``operator valued weight'' \cite{Haagerup:1979ovwI,Haagerup:1979ovwII}, which is the unique\footnote{More precisely, $\mathcal{F}_\lambda$ is unique (up to a choice of $\lambda$) because it is also normal, faithful, and semi-finite: see \cite{Balasubramanian:2026ymu} for more details.} map from  density matrices for $\mathcal{A}_a$ to density matrices for $\mathcal{C}_a$, up to the choice of positive function $\lambda(R)$ which defines $\mathcal{F}_\lambda$. This freedom in choosing $\lambda$ is fixed by demanding that $\mathcal{F}_\lambda$ sends normalized states on $\mathcal{A}_a$ to normalized states on $\mathcal{C}_a$, which in turn requires a notion of trace on $\mathcal{C}_a$ with respect to which $\mathcal{F}(\rho_a)$ is normalized.

However, the physical trace \eqref{eq:physicaltrace} assigns an infinite trace to $\mathcal{C}$ by itself: each factor $\Id_{\Ha_a(R)}$ within \eqref{eq:centralop} is the identity on an infinite dimensional Hilbert space, and so the trace $\Tr_a[\mathcal{C}]$ diverges universally within each sector $R$. Therefore, we must renormalize the trace one more time to define a trace $\Tr_{\mathcal{C}_a}$ for the central operators. For any choice of measure $d\lambda(R) \equiv \lambda(R)^{-1} d\mu(R)$ over the representations, we can define a trace
\begin{align}
    \Tr^\lambda_{\mathcal{C}_a}[\mathcal{C}] = \int d\lambda(R) \,C(R)
\end{align}
for the central operators. With respect to this notion of trace, we can see that
\begin{align}
    \Tr^\lambda_{\mathcal{C}_a}[\mathcal{F}_\lambda(\rho_a)] = \int d\lambda(R) \lambda(R) q_R = \int d\mu(R) \,q_R = 1 \,.
\end{align}
So, the trace $\Tr^\lambda_{\mathcal{C}_a}$ uniquely specifies the map $\mathcal{F}_\lambda$ we should use to isolate the state on the center of the algebra. For any pair of compatible $(\mathcal{F}_\lambda(\rho_a) , \Tr^\lambda_{\mathcal{C}_a})$, the expectation value of any state independent operator $\mathcal{C}$ will be independent of $\lambda$. However, because the entropy due to the center can be thought of as the expectation value of the state-dependent operator $-\ln(\mathcal{F}_\lambda(\rho_a))$, we must still fix the function $\lambda$ appearing in the trace.

For the algebra $\mathcal{A}_a$, the measure we used to define the trace was the Plancherel measure $d\mu(R)$. 
However, in \cite{Balasubramanian:2026ymu,Balasubramanian:2026vlr}, we showed that for the center $\mathcal{C}_a$, the preferred trace is instead defined using the \emph{microcanonical measure} $dR$.\footnote{This measure is preferred because the center $\mathcal{C}_a$ does not depend on the entire gauge group, but only its conjugacy classes. The microcanonical measure $dR$ is the Fourier dual of the group-basis measure that treats the conjugacy classes of the gauge group uniformly. In contrast, the entire algebra $\mathcal{A}_a$ does depend on the entire gauge group, so we define its trace using the Fourier dual of the Haar measure, i.e., the Plancherel measure.}  Whereas the Plancherel measure weights the representations $R$ non-uniformly to maintain the unitarity of the Fourier transform, the microcanonical measure weights the representations $R$ uniformly. When $G$ is a compact Lie group, the microcanonical measure is the counting measure over the discrete set of representations. When $G$ is a non-compact Lie group, the microcanonical measure is instead the Fourier dual of the disjoint union of Haar measures over each Cartan subgroup of $G$. In 3D gravity in AdS spacetimes, the microcanonical measure is the Lebesgue measure over the Liouville momenta $P$ (equivalently, the uniform measure over the lengths $\ell$ of geodesic curves on the Cauchy slice). We refer the reader to \cite{Balasubramanian:2026ymu,Balasubramanian:2026vlr} for more details about the microcanonical measure, and will use this measure to define the trace of the central operators. 

Thus, defining the trace over the center with the microcanonical measure, we can see that the map $\mathcal{F}$ which isolates the component of the state $\rho_a$ which acts on the center as
\begin{align}
    \mathcal{F}(\rho_a) = \int d\mu(R) (\mu(R) q_R) \Id_{\Ha_a(R)} \otimes \Pi_R \,. \label{eq:OVW}
\end{align}
The extra factor $\mu(R)$ is defined as the ratio between the Plancherel and microcanonical measures
\begin{align}
    \frac{d\mu(R)}{dR} \equiv \mu(R)\,. \label{eq:muRdef}
\end{align}
This factor will appear in the decomposition of the entropy of $\rho_a$, as we will now see.

\subsection{The entropy}\label{sec:entropy}

We now use the replica trick to compute the entropy $S(\rho_a)$. Explicitly using the projector property of $\Pi_R$ \eqref{eq:Piproj} and the definition of the physical trace \eqref{eq:physicaltrace}, we see that
\begin{align}
    S(\rho_a) &= - \partial_n\Tr_a\left[\rho_a^n\right]_{n=1}
    \\&= - \partial_n \int d\mu(R) q_R^n \tr_{\Ha_a(R)}[\rho_R^n] \,,
    \\&= \int d\mu(R) \left[-q_R \ln(q_R) - q_R \tr_{\Ha_a(R)}[\rho_R \ln(\rho_R)] \right]\,. \label{eqn:entropy_first}
\end{align}
The first term captures the uncertainty from the distribution over representations $R$. However, we can further decompose this term by noting that the entropy due to the center is given by
\begin{align}
    S(\mathcal{F}(\rho_a)) &=  -\Tr_{\mathcal{C}_a}[\mathcal{F}(\rho_a) \ln(\mathcal{F}(\rho_a))] \,,
    \\&= - \int dR \,p_R \ln(p_R)  \,.
\end{align}
where $p_R = \mu(R) q_R$.
Using this expression, \eqref{eqn:entropy_first} becomes
\begin{align}
    S(\rho_a) &= S(\mathcal{F}(\rho_a)) + \langle \hat{A} \rangle_\rho + S_{\mathrm{bulk}}(\rho_a)\,. \label{eq:entropyformula}
\end{align}
where $\langle \hat{A} \rangle_\rho$ is the expectation value of the operator we define shortly in \eqref{eq:areaop}. We will now explain each of these three terms in more detail, and refer the reader to \cite{Balasubramanian:2026ymu,Balasubramanian:2026vlr} for more details in the case of 3D gravity in AdS spacetimes (and other related TQFTs).

\subsubsection{The entropy due to the center} \label{sec:centerentropy}

The first term $S(\mathcal{F}(\rho_a))$ is the contribution to $S(\rho_a)$ which arises from the correlations due to operators $\mathcal{C}$ which lie on the corner $\partial a$ between the region and its complement. Whenever there is a continuum of representations $R$, this contribution to the entropy is not bounded below. This is because we can always take $p_R$ to be arbitrarily sharply peaked, so there is no minimum entropy state on the center $\mathcal{C}_a$. 

Because the corner symmetry group is non-compact, any UV deformation of it (as discussed in Sec.~\ref{sec:em_grav} ) which preserves the structure of this group above the cutoff will have a continuum of representations in the limit the cutoff is removed. However, any deformation which adds an underlying discreteness (which is a recurring theme in gravity \cite{Saad:2018bqo,Penington:2019npb,Almheiri:2019psf,Penington:2019kki,Almheiri:2019qdq,Iliesiu:2024cnh,Akers:2025ynh,Balasubramanian:2022lnw,Balasubramanian:2022gmo,Chandra:2022fwi,Balasubramanian:2024lqk,Balasubramanian:2025hns,Balasubramanian:2026azk}) will render this term positive definite. In order for the effective theory we are describing here to be a good approximation to all orders in perturbation theory, the discretization scale of the representations must be exponentially small in $G_N$. In other words, we expect non-perturbative effects to render the spectrum of representations discrete, with an average level spacing which is exponentially small in $G_N$, so that $S(\mathcal{F}(\rho_a))$ is bounded below.

\subsubsection{The area operator} \label{sec:areaop}

The second term $\langle \hat{A} \rangle_\rho$ is the expectation value of the state independent operator $\hat{A}$, defined as
\begin{align}
    \hat{A} = \int d\mu(R) \ln(\mu(R)) \cdot \Id_{\Ha_a(R)} \otimes \Pi_R \,.\label{eq:areaop}
\end{align}
The expectation value $\langle \hat{A} \rangle_\rho$ is taken in the state $\rho_a$. This term arises because of the difference in traces for the algebra $\mathcal{A}_a$ and its center $\mathcal{C}_a$. Equivalently, this term arises because the Plancherel measure $d\mu(R)$ does not weigh the representations $R$ uniformly. If we think of $\mu(R)$ as a density of states for the representations $R$, in the sense that $d\mu(R) = \mu(R) dR$ in the same way that $dN = \rho(E) dE$ in statistical mechanics, then this contribution to the entropy can be thought of as being analogous to the Boltzmann entropy $\ln(\rho(E))$.

To see that $\hat{A}$ should be specifically identified with the area operator, and not some other observable, consider the case of 3D gravity in AdS spacetimes. Here, we briefly review the argument laid out in \cite{Balasubramanian:2026ymu,Balasubramanian:2026vlr}, and refer the reader there for more details.
In the first order formalism of gravity, and on a fixed background topology $\mathcal{M}$ for the spacetime which admits an on-shell solution, the action of 3D gravity is the same as that of $\SL(2,\R) \times \SL(2,\R)$ Chern--Simons theory \cite{Witten:1988hc}.\footnote{More precisely, a $\SL(2,\R) \times \SL(2,\R)$ gauge transformation of the first order variables is equivalent to a diffeomorphism, up to a term proportional to the equations of motion.} In these variables, the gauge group $\SL(2,\R) \times \SL(2,\R)$ plays the role of the diffeomorphism group, as it is the gauge group of the theory. To account for the invertibility of the spacetime metric, which the Chern--Simons description neglects, one can instead consider two copies of Virasoro TQFT (VTQFT) \cite{Collier_2023,Collier:2024mgv},\footnote{Equivalently, one can use conformal Turaev--Viro (CTV) theory \cite{Hartman:2025cyj,Hartman:2025ula}. See also \cite{Jafferis:2025vyp,Jafferis:2025jle,Jafferis:2025yxt} for another model which may account for the sum over topologies that the VTQFT/CTV description of 3D gravity neglects, as well as \cite{Mertens:2022ujr,Mertens:2025ydx,Chen:2024unp} for other related models.} which can be thought of as a refinement of the Chern--Simons description of 3D gravity which accounts for this invertibility. At the same time, the role of the group $\SL(2,\R)$ is replaced with a mathematical structure known as the modular double of the quantum group $\mathcal{U}_q(\sl(2,\R))$, which can be thought of as a deformed version of the Lie algebra of $\SL(2,\R)$ \cite{Faddeev:1999fe,Teschner:2012em,Mertens:2022aou,Bytsko:2002br}.\footnote{$q$ is a numerical deformation parameter which takes the value $q = \exp(i\pi \, 4 G_N)$, in units where the AdS radius has been set to $1$, and $G_N$ is Newton's constant. The modular doubling appends another copy of this structure with $\widetilde{q} = \exp(i\pi (4 G_N)^{-1})$.}

In this case, the analog of the representations $R$ are given by a pair of positive real numbers $(P_+, P_-)$ called the Liouville momenta \cite{Teschner:2003em}. If we rescale these Liouville momenta as
\begin{align}
    P_\pm = \frac{\ell_\pm}{8 \pi \sqrt{G_N}}\,,
\end{align}
where $G_N$ is Newton's constant, then in the $G_N \to 0$ limit with $\ell_\pm$ held fixed, it is known \cite{Carlip:1994gy} that the eigenvalues of the area operator are given by
\begin{align}
    \ln(\mu(P_+, P_-)) \approx \frac{\ell}{4 G_N} \,, && \ell = \frac{\ell_+ + \ell_-}{2} \,.
\end{align}
When the corner $\partial a$ is homologous to the horizon of a rotating BTZ black hole, $\ell$ is the area of the outer horizon of the black hole \cite{Carlip:1994gc}. When the corner $\partial a$ is homologous to a boundary subregion of a global AdS spacetime, then $\ell$ is the length of the boundary anchored geodesic \cite{Balasubramanian:2026vlr,Wong:2022eiu}. So in this case, $\langle \hat{A} \rangle_\rho$ really does measure the area of a co-dimension 2 surface in the spacetime. We propose that this continues to hold for more general theories of gravity.

\subsubsection{The bulk entropy} \label{sec:bulkent}

Finally, the third term $S_{\mathrm{bulk}}(\rho_a)$ is the average entropy within each sector
\begin{align}
    S_{\mathrm{bulk}}(\rho_a) &= \int d\mu(R) q_R S(\rho_R) \,,\\
    S(\rho_R) &= - \tr_{\Ha_a(R)}[\rho_R \ln(\rho_R)] \,.
\end{align}
We call this the \emph{bulk entropy}, because unlike the other two terms, which descend from operators with support on the corner itself, the bulk entropy depends on details of the state in the bulk of the subregion $a$.

\subsubsection{Further comments}

\paragraph{Quantum inequalities.} One important point is that $S(\rho_a)$ really is a von Neumann entropy, so inequalities satisfied by arbitrary quantum states, such as strong subadditivity, are automatically satisfied.\footnote{Strictly speaking, because $S(\rho_a)$ is only defined up to a state-independent constant because of the continuum of representations, this is only guaranteed for \emph{balanced} inequalities where the number of terms on the RHS and LHS of the inequality are equal \cite{Hubeny:2018ijt}, so that this constant cancels from both sides of the inequality.} But holographic entropy is subject to stronger constraints than generic quantum states. These strictly holographic constraints on holographic entropy are called the facets of the holographic entropy cone \cite{Bao:2015bfa}. These constraints follow directly from the geometrization of entropy in the bulk in time symmetric states.
For the moment, focus on $3$ spacetime dimensions. In the ground states of gapped QFTs, there is a contribution to the entropy analogous to that from $\hat{A}$ called the topological entanglement entropy \cite{Kitaev:2005dm,Levin:2006arx}. We showed in \cite{Balasubramanian:2026ymu,Balasubramanian:2026vlr} that for a toy model of 3D gravity in AdS spacetimes, where the metric is not constrained to be invertible, the area contribution to the entropy can be thought of as a generalization of the topological entanglement entropy.\footnote{As explained in \cite{Balasubramanian:2026ymu,Balasubramanian:2026vlr}, we expect that the invertibility of the metric can be accounted for by replacing the classical group representation theory data with an analogous set of data from the modular double of the quantum group $\mathcal{U}_q(\sl(2,\R))$.} It was shown in \cite{Naskar:2024mzi} that for finite gauge groups (or compact gauge groups at finite level), topological entanglement entropy in 3D satisfies all facet inequalities of the holographic entropy cone. Thus, assuming an analogous result holds true for the more complicated gauge group $\mathrm{ECS}$, this could constitute a proof on the quantum side that the entropy from the area operator \eqref{eq:areaop} also satisfies all facets of the holographic entropy cone in 3D. It would be very interesting to extend these results to the gauge groups more relevant for the gravitational case, or to higher dimensions.

\paragraph{Quantum error correction.} The structure of the reduced state in \eqref{eq:reducedstate} is not only necessary for gauge invariance of the global state. It is also consistent with the interpretation of the holographic map as a quantum error correction protocol \cite{Harlow:2016vwg}.\footnote{As discussed in \cite{Balasubramanian:2026ymu} and reviewed in Sec.~\ref{sec:centerentropy}, this interpretation requires non-perturbative effects which render the area spectrum discrete, with an average level spacing which is exponentially small in $G_N$, so that $S(\rho_a)$ is bounded below.}
The fact that the edge mode state $\Pi_R$ is maximally mixed within each superselection sector $R$ is expected from AdS/CFT \cite{Akers:2018fow,Dong:2018seb,Harlow:2016vwg} and plays a crucial role in the quantum error correction underlying complementary recovery. We could also view the maximal entropy within each superselection sector as a refinement of Jacobson's entanglement equilibrium hypothesis, which posits that the semi-classical vacuum is the maximal entropy state \cite{Jacobson_2016}. For sufficiently symmetric regions $a$, this condition is sufficient to recover the semi-classical Einstein equations.

\paragraph{Observers and edge modes.} Recent investigations into quantum gravity have suggested the necessity to include additional degrees of freedom to define a subsystem in gravity \cite{Balasubramanian:2026ymu,Balasubramanian:2026vlr,Chandrasekaran:2022eqq,Witten:2023qsv,Witten:2023xze,Chen:2024rpx,AliAhmad:2023etg,Faulkner:2024gst,DeVuyst:2024khu,Fewster:2024pur,Abdalla:2025gzn,Harlow:2025pvj,Soni:2023fke,Akers:2024wab,Klinger:2023tgi,Klinger:2023auu,Krishnan:2023fnt,Gomez:2022eui,Ciambelli:2021nmv,Balasubramanian:2023dpj,Klinger:2026tws,Leutheusser:2022bgi,Witten:2021unn,Chandrasekaran:2022cip,Jensen:2023yxy,Kudler-Flam:2023qfl}. Often, these degrees of freedom are called ``observers'', ``edge modes'', or ``quantum reference frames''. In our case, the edge modes introduced by gravity arose from studying the diffeomorphism constraints near the corner $\partial a$ where we factorized the Hilbert space, so in that sense they are ``internal'' to the theory, despite being ``additional'' degrees of freedom. It would be interesting to make the connection between all of these approaches more precise.

\section{The path integral: bulk--then--edge} \label{sec:path_int}

In the previous section, we used canonical quantization to argue for the entropy formula \eqref{eq:entropyformula}. We will now use the path integral to argue that \eqref{eq:entropyformula} is indeed the generalized entropy, or more precisely, its generalization to finite regions $a$.

\subsection{The path integral}

The entropy formula \eqref{eq:entropyformula} required a regulator for $\mathrm{ECS}$ which renders the gauge group locally compact. To see what happens when we remove this regulator, it is useful to rewrite the integral over representations into an integral over the embedding fields themselves. This is equivalent because the representations $R$ and the embedding fields $\phi$ (which are represented by a particular group element $g \in \mathrm{ECS}$) are Fourier dual to each other. In particular, the Plancherel measure over representations is dual to the Haar measure over group elements, and in the limit that we remove the regulator, the Haar measure over embedding fields becomes the ultralocal path integral measure over those fields.

To see this, consider the path integral of a free scalar field $\varphi$. Intuitively, the ultralocal path integral measure $D\varphi$ is a Haar measure for the group of field space translations $\varphi \to \varphi + \delta \varphi$. The reason that $D\varphi$ does not exist at the level of rigor suitable for mathematicians is that this group is not locally compact. When we UV regulate the theory by placing it on a lattice $L$ with lattice spacing $\Lambda^{-1}$, the group of field space translations becomes locally compact, and the measure is defined as $D\varphi_\Lambda \sim \prod_{x \in L} d\varphi_x$. We then define the continuum measure $D\varphi$ as the $\Lambda \to \infty$ limit of $D\varphi_\Lambda$.

In our setting, the integral over $\mathrm{ECS}$ representations in the definition of the reduced state \eqref{eq:reducedstate} indicates that the embedding fields $\phi: a \into \Sigma$ will be part of the definition of the path integral, because $\phi$ is Fourier dual to the representations $R$. Thus, as we remove the regulator which deforms $\mathrm{ECS}$, the path integral we need to compute will take the schematic form
\begin{align}
    \rho_a = \int \frac{D\psi D\phi}{\mathrm{Diff}} e^{-S[\psi,\phi]} \,,
\end{align}
for some action $S[\psi,\phi]$ we will determine shortly. Here, $\psi$ is a collective notation for all bulk fields, including the metric. As usual, the boundary conditions for the fields $(\psi,\phi)$ determine the state $\rho_a$. $\mathrm{Diff}$ denotes the volume of the diffeomorphism group, which acts on the bulk fields $\psi$ and the embedding map $\phi$.

For concreteness, let $\ket{0}$ be the Hartle--Hawking state of the gravitational theory specified by the Lagrangian $L[\psi]$ for the bulk fields. After accounting for the universal divergence associated with the non-compactness of the diffeomorphism group (see \cite{Balasubramanian:2025rcr,Balasubramanian:2026ymu,Balasubramanian:2026vlr} for more details in the case of 3D gravity), the factorization map $V: \Ha_\Sigma \to \widetilde{\Ha}_a \otimes \widetilde{\Ha}_{a'}$ is an isometry. Therefore, the norms $\braket{0}$ and $\bra{0} V^\dagger V \ket{0}$ should agree. In the path integral language, this means that we should find that
\begin{align}
    \mathcal{Z}(a) \equiv \Tr_a[\rho_a] = \braket{0} \,. \label{eq:Zadef}
\end{align}
The norm $\braket{0}$ is the usual path integral of the Hartle--Hawking state $\ket{0}$ of the fields $\psi$, without any edge modes. This equality constrains the possible action $S[\psi,\phi]$ of the factorized state. 

The action of the path integral preparing the unfactorized state $\ket{0}$ is given by
\begin{align}
    S[\psi] = \int^\Sigma_{M_-} L[\psi] \,.\label{eq:HHaction}
\end{align}
Here, $L[\psi]$ is the Lagrangian density of the theory, thought of as a differential form so it includes the spacetime integration measure. $M_-$ denotes the usual Euclidean manifold preparing the state $\ket{0}$, such as the half-sphere for the dS vacuum, or the half disk for the AdS vacuum. Similarly, we write $M_+$ for its time reverse, which computes $\bra{0}$. Furthermore, $\psi$ denotes the field configuration of the fields (including the metric) within $M_-$. The fields at $\Sigma$ (as well as any other boundary components of $M_-$) are held fixed, and are indicated here by the superscript on the integral. The value of the path integral over all fields $\psi$, as a functional of these boundary fields at $\Sigma$, then defines the wave functional for the state $\ket{0}$.

We propose that the action of the path integral preparing the factorized state $V\ket{0}$ is given by
\begin{align}
    S[\psi,\phi] = \int_{M_-}^{\Sigma \setminus \phi(\partial a)} L[\psi] = \int_{M_-}^{a \sqcup a'} \phi^* L[\psi] \,. \label{eq:VHHaction}
\end{align}
The notations in \eqref{eq:VHHaction} are mostly the same as in \eqref{eq:HHaction}, with the following modifications. First, we fix an embedding $\phi: a \into \Sigma$, which defines a subregion of the Cauchy slice (and similarly for $\phi':a' \into \Sigma$, as above). Instead of holding the field values fixed on the Cauchy slice $\Sigma$, we hold the fields within $\phi(a)$ and $\phi'(a')$ fixed, thought of as an embedded submanifold $\phi(a) \sqcup \phi'(a') \equiv \Sigma \setminus \phi(\partial a) \subset \partial M_-$.\footnote{The symbol $\sqcup$ indicates that $\Sigma \setminus \phi(\partial a)$ is the disjoint union of $\phi(a)$, $\phi'(a')$: the corner is treated independently within each subregion.}
This can be thought of as the same Cauchy slice $\Sigma$ as in the unfactorized case, but with a neighborhood of the corner $\phi(\partial a)$ removed from the spacetime.
Equivalently, we can define the wave functional of $V\ket{0}$ by fixing the field configurations within the ``model subregions'' $a,a'$ after using the pullback $\phi^*$ to translate the field values $\psi(\phi(x))|_{\phi(a)} = (\phi^*\psi)(x)|_a$. This is the perspective taken in the RHS of \eqref{eq:VHHaction}. For any boundary components of $\Sigma \setminus \phi(\partial a)$ which descend from boundary components of $\Sigma$, the fields $\psi$ are given the same boundary conditions as in the unfactorized case. But because $\Sigma \setminus \phi(\partial a)$ has an additional boundary component at the corner (really, two: one for $a$ and one for $a'$), we must additionally specify the boundary condition for the fields $\psi$ at the corner for this path integral to be well-defined, which we now explain.

Recall that $a,a'$ are \emph{open} subregions of $\Sigma$. Thus, $\phi(a) \sqcup \phi'(a') \equiv \Sigma \setminus \phi(\partial a)$ does not include the corner itself. In the unfactorized state, there is only a single corner $\phi(\partial a)$ within $\Sigma$. In the factorized state, however, the corner has been effectively ``doubled'': the closures of $\phi(a)$ and $\phi'(a')$ each have a boundary component which descends from the corner. As explained above, the condition that $V\ket{0}$ descends from the gauge-invariant, unfactorized state $\ket{0}$ implies that the field values assigned to each of these corners must match, and so we must impose boundary conditions at this corner which ensure this holds. The standard way to implement this matching across $a$ and $a'$ is with the so-called ``shrinkable'' boundary conditions \cite{OhmoriTachikawa2014Entangling,DonnellyWong2016EBrane,Donnelly:2018ppr,Ball:2024hqe,Jafferis:2019wkd,Mertens:2022ujr,Wong:2022eiu,Jiang:2020cqo}, which we define below when we compute the norm of $V\ket{0}$.\footnote{In principle, we should also sum over all topologies of $M_-$ consistent with the boundary conditions on $\partial M_-$ (the field values on $\Sigma$ and any asymptotic boundaries), as well as over topologies of $\Sigma$ itself. We neglect these non-perturbative effects in this paper, and assume they provide subdominant contributions. In particular, $\Sigma$ is the Cauchy slice of the dominant saddle, which is how its topology is determined.} But first, we will explain the role of the boundary condition at the corner more intuitively.

In principle, there are two notions of ``matching'' that this boundary condition should implement. First, the ``UV matching'' which ensures that the field values of $\psi$ do not jump discontinuously across the corner. This condition is necessary regardless of the gauge symmetry, and arises because the path integral measure $D\psi$ is defined over continuous (possibly non-differentiable) fields. Second is the ``IR matching'' that ensures that the fields at the corner transform in the same representation of the gauge group. This ensures that the factorization map $V$ does not introduce a spurious charge at the corner when the state is glued back to the unfactorized state: see the discussion of ``defect-free'' factorization maps in \cite{Balasubramanian:2026ymu} for more details in the case of 3D gravity (and related topological models).

Finally, we note that the second expression in \eqref{eq:VHHaction} makes the gauge invariance of this path integral manifest. There, $\phi^* L$ denotes the pullback of the fields $\psi$ on $\Sigma$ along the embedding $\phi$, so that they become fields on the fixed model regions $a$ and $a'$. A diffeomorphism acting on the fields on $\Sigma$ also transforms $\phi$, and leaves the pulled back fields invariant, even though the diffeomorphism need not vanish on the corner. Thus, as expected, diffeomorphisms are a gauge symmetry of this action, and must be quotiented out of the field space.

When we glue two copies of the state $V\ket{0}$ together to compute its norm, the total action of the glued path integral will be
\begin{align}
    S[\psi,\phi] = \int_{M_- \cup M_+ \setminus \phi(\partial a)} \phi^* L[\psi] \,.
\end{align}
Here, we identify the fields within $a_+$ and $a_-$ (and similarly for $a_+'$ and $a_-'$). We then integrate over all possible field values, subject to these identifications, on $\Sigma$, effectively gluing them together.\footnote{The symbol $\cup$ indicates that $M_-,M_+$ have been glued along their common boundary: the remaining $\setminus \phi(\partial a)$ indicates that we have excised a neighborhood of the corner $\phi(\partial a)$ from the resulting manifold, which we discuss below.} Furthermore, because we quotient out by diffeomorphisms in the measure of the path integral, gauge equivalent embedding maps should be identified in this path integral, and only one embedding map per gauge orbit should be included. 
However, because each region $a,a'$ is an \emph{open} subregion of $\Sigma$, the corner is excluded from the manifold: it has been cut out of $\Sigma$, and hence is not part of the glued manifold preparing $\bra{0} V^\dagger V \ket{0}$. 
This leaves behind a co-dimension 2 defect within the glued manifold. We denote the resulting manifold as $M_- \cup M_+ \setminus \phi(\partial a)$. The behavior of the fields near the defect $\phi(\partial a)$ is dictated by the shrinkable boundary conditions we impose at the corner, which we now explain.

Instead of excising the corner from $M_- \cup M_+$, which is a codimension 2 object, we can excise a neighborhood of the corner $N_\epsilon$. By $N_\epsilon$, we really mean a family of neighborhoods of $\phi(\partial a)$ labeled by a parameter $\epsilon$, such that $N_{\epsilon} \subset N_{\epsilon'}$ for $\epsilon < \epsilon'$. Furthermore, we demand that $\phi(\partial a) = N_0 \equiv \lim_{\epsilon\to 0} N_\epsilon$, so that in the limit $\epsilon \to 0$, we recover the topology we started with. In other words, in the limit $\epsilon \to 0$, we have effectively glued back in the corner.
But for fixed $\epsilon$, we must impose boundary conditions on $\partial N_\epsilon$ for the variational principle to be well-defined. We impose a family of ``shrinkable'' boundary conditions on $N_\epsilon$  \cite{OhmoriTachikawa2014Entangling,DonnellyWong2016EBrane,Donnelly:2018ppr,Ball:2024hqe}. Shrinkable boundary conditions are defined such that, as a function of $\epsilon$, any correlation function behaves as if there was no excision at all in the limit $\epsilon \to 0$. In other words, we demand that
\begin{align}
    \lim_{\epsilon \to 0} \langle \cdots \rangle_{M_- \cup M_+ \setminus N_\epsilon} = \langle \cdots \rangle_{M_+ \cup M_-} \,. \label{eq:shrinkable}
\end{align}
Shrinkable boundary conditions will generally involve $\epsilon$ dependent counterterms which cancel divergences that can arise from the naive $\epsilon\to 0$ limit. See e.g. \cite{OhmoriTachikawa2014Entangling,DonnellyWong2016EBrane,Donnelly:2018ppr,Ball:2024hqe} for more details about shrinkable boundary conditions.

Because of the shrinkable boundary conditions, after integrating out the embedding fields we are left with the usual path integral, with action
\begin{align}
    S[\psi] = \int_{M_1}L[\psi] \,,
\end{align}
where $M_1 = M_- \cup M_+$ is the manifold resulting from gluing $M_-$ and $M_+$ along their common boundary $\Sigma$. This follows from the quotient by simultaneous diffeomorphisms of $\phi$ and $\psi$ on $\Sigma$: each physical field configuration on the cut appears only once, so integrating over the field values on $\Sigma$ simply glues the two manifolds together. This follows from the shrinkable boundary conditions \eqref{eq:shrinkable}. Also, note that this is precisely the action of the path integral computing $\braket{0}$. So at least formally, the shrinkable boundary conditions reproduce the fact that $V$ is an isometry in the path integral language.\footnote{Implicit was a cancellation between the path integral over embedding maps $\phi$ and part of the volume of the gauge group corresponding to simultaneous diffeomorphisms of $\phi$ and the bulk fields $\psi$. While we do not do so in this paper, it would be interesting to use the more rigorous techniques of BRST-BV quantization to check this argument more carefully.
} 

One might wonder what we gain by introducing the embedding map $\phi$, given that it drops out of the norm $\bra{0} V^\dagger V \ket{0}$. Its role becomes clear when we construct the reduced state $\rho_a$. In a quantum field theory without gauge symmetry, $\rho_a$ is computed by preparing two copies of the state, gluing them by integrating over the fields on $a'$, and leaving the fields on $a$ unintegrated. In gravity, the coordinate location of the cut $\partial a$ is not gauge invariant, so this construction does not immediately apply. However, the embedded cut $\phi(\partial a)$ \emph{is} diffeomorphism invariant, because $\phi$ transforms along with the fields. We can therefore define $\rho_a$ by gluing the fields on $\phi'(a')$, and keeping the fields on $\phi(a_\pm)$ fixed, where $a_\pm$ is the subregion of $\partial M_\pm$ that we are interested in. 
The action for the path integral computing $\rho_a$ is then
\begin{align}
    S[\psi,\phi] = \int_{M_1 \setminus \phi(a)}^{\phi(a_+) \sqcup \phi(a_-)} L[\psi] = \int_{M_1 \setminus \phi(a)}^{a_+ \sqcup a_-} \phi^* L[\psi]\label{eq:rhoaaction}
\end{align}
Here, $M_1 \setminus \phi(a)$ denotes $M_- \sqcup M_+$ glued along the two copies of $\phi'(a')$, as shown in Fig.~\ref{fig:rhoa}. Equivalently, this is $M_1$ with a neighborhood of the embedded subregion $\phi(a)$ removed. This is well-defined because a diffeomorphism of the fields within $a$, together with the corresponding action on $\phi$, leaves $\phi(a)$ invariant. We continue to impose shrinkable boundary conditions near the corner $\phi(\partial a)$.

\begin{figure}
    \centering
    \begin{tikzpicture}[scale=2]
        \def\xx{0.3}
        \def\eps{0.025}
        \draw[thick] (0,0) circle (1);
        \draw[thick,dashed] (-1,0) -- (-\xx,0) (\xx,0) -- (1,0);
        \draw[dashed,thick] (-\xx,\eps) -- (\xx,\eps) (-\xx,-\eps) -- (\xx,-\eps);
        \filldraw[fill=white,thick] (-\xx,0) circle (0.05);
        \filldraw[fill=white,thick] (\xx,0) circle (0.05);
        
        \draw[thick,->] (-0.5-\xx/2,-0.25)-- (-0.5-\xx/2, -0.05) node[anchor=north] at (-0.5-\xx/2, -0.25) {$a'$};
        \draw[thick,->] (0.5+\xx/2,-0.25)-- (0.5+\xx/2, -0.05) node[anchor=north] at (0.5+\xx/2, -0.25) {$a'$};
        \draw[thick,->] (0,-0.25)-- (0, -0.05) node[anchor=north] at (0, -0.25) {$a_-$};
        \draw[thick,->] (0,0.25)-- (0, 0.05) node[anchor=south] at (0, 0.25) {$a_+$};
        \draw[thick,->] (-\xx-0.25,0.25) -- (-\xx-0.05,0.05) node[anchor=south] at (-\xx-0.25,0.25) {$\phi(\partial a)$};
        \draw[thick,->] (\xx+0.25,0.25) -- (\xx+0.05,0.05) node[anchor=south ] at (\xx+0.25,0.25) {$\phi(\partial a)$};
        \draw[thick,->] (-1.25,0) -- (-1-0.05,0) node[anchor=east] at (-1.25,0) {$\Sigma$};
    \end{tikzpicture}
    \caption{An example of our path integral proposal when $\ket{0}$ is prepared using the Euclidean disk. For notational simplicity, we have written e.g. $a'$ for $\phi'(a')$. To compute $\rho_a$, one is instructed to glue the fields on the slice $\Sigma$ within the partial slice $\phi'(a')$ (which is a diffeomorphism invariant specification), and leave the fields within $\phi(a)$ fixed. This is the usual prescription in quantum field theory: the major difference is that the gauge invariant specification of the cut $\phi(\partial a)$ requires the auxiliary embedding field $\phi$. This definition requires us to excise the corner $\phi(\partial a)$ from the manifold and impose ``shrinkable boundary conditions'', which is equivalent to the corner being undetectable in the limit we remove the cutoff.
    }
    \label{fig:rhoa}
\end{figure}

Let us check that the path integral with action \eqref{eq:rhoaaction} really does produce an operator $\rho_a$ on the extended Hilbert space $\widetilde{\Ha}_a$. To do so, note that $\widetilde{\Ha}_a$ supports a faithful unitary action of $\mathrm{ECS}$. These unitaries are generated by Hermitian operators which satisfy the symmetry algebra 
\begin{align}
    [H_\eta, H_\xi] = -i\,H_{[\eta,\xi]}\,, \label{eq:charges}
\end{align}
where $[\eta,\xi]$ denotes the Lie bracket of the two vector fields $\eta,\xi$, thought of as generators of $\mathrm{ECS}$. Thus, the Hilbert space defined by path integral with action \eqref{eq:rhoaaction} must support the existence of such operators.

In \cite{Ciambelli:2021nmv}, the  conditions necessary for the existence of such charges were investigated using covariant phase space techniques. The authors of \cite{Ciambelli:2021nmv} treated the embedding map $\phi$ as a dynamical field, and analyzed  possible boundary conditions at $\partial a$ which allow for a faithful representation of ECS, i.e., when \eqref{eq:charges} holds. They showed that \emph{precisely} because the embedding map is treated as a dynamical field, it contributes to the variation of the action, and so the Hamiltonian charges that generate $\mathrm{ECS}$ indeed form the required canonical commutation relations to generate a unitary representation of $\mathrm{ECS}$. Furthermore, the symplectic structure of \eqref{eq:rhoaaction} generates the maximal such set of non-zero charges. Thus, because the extended Hilbert space $\widetilde{\Ha}_a$ also supports the maximal, faithful representation of $\mathrm{ECS}$ (essentially by definition), the path integral prescription of \eqref{eq:rhoaaction} should prepare a state of $\widetilde{\Ha}_a$.

Crucially, the charges generating these $\mathrm{ECS}$ unitaries satisfy the necessary canonical commutation relations regardless of the boundary conditions of the fields (including the metric) at the corner $\partial a$ \cite{Ciambelli:2021nmv}. In other words, we do \emph{not} need to impose any condition such as extremality of the area of the corner at $\partial a$ to get a valid state in the extended Hilbert space. 

\subsection{R\'enyi entropy}

We now compute the R\'enyi entropies of $\rho_a$ in the $G_N\to 0$ limit. As in Sec.~\ref{sec:edge}, we take the Cauchy slice $\Sigma$ to be a moment of time symmetry of $M_1$, which defines the Cauchy slice that $a$ lives on. The computation parallels the canonical analysis of Sec.~\ref{sec:entropy}, and its purpose is to identify where the terms of \eqref{eq:entropyformula} arise from the perspective of the gravitational path integral. 

Recall from \eqref{eq:Zadef} that $\mathcal{Z}(a) = \Tr_a[\rho_a] = \braket{0}$, because the embedding field drops out of the calculation of the norm. For the replicated path integral, we similarly write
\begin{align}
    \mathcal{Z}(a^n) = \Tr_a[\rho_a^n] \,,
\end{align}
so that the von Neumann entropy is
\begin{align}
    S(\rho_a) = -\partial_n \left(\frac{\mathcal{Z}(a^n)}{\mathcal{Z}(a)^n}\right)_{n=1}\,.
\end{align}
 
We will now compute $\mathcal{Z}(a^n)$ using the replica trick.
Consider $n$ copies of $\rho_a$. In principle, each path integral contains multiple contributions of different bulk topologies which must be summed over. As above, we assume that each $\rho_a$ calculation is dominated by a single bulk topology $M_1$. We then cyclically glue the $i$th copy of $a_-$ to the  $(i+1)$th copy of $a_+$. This prepares a new topology $M_n$ on which we compute the path integral. Holding fixed the representation $R$ of the fields at the corner, we then compute the path integral around this background. Finally, we integrate over all possible representations $R$. 

The full path integral includes a sum over all bulk topologies compatible with the replicated boundary conditions and a single embedded corner $\phi(\partial a)$, and the dominant topology need not be the $n$-fold cover of $M_1$. This is, for example, the mechanism that leads to the Page curve for evaporating black holes \cite{Almheiri:2019psf,Penington:2019npb,Penington:2019kki,Almheiri:2019hni}. We restrict to states $\rho_a$ for which the $n$-fold cover of $M_1$ dominates, i.e., we neglect replica wormhole contributions to $\mathcal{Z}(a^n)$. For boundary subregions in AdS/CFT, see \cite{Akers:2020pmf} for work on when this approximation breaks down. It would be interesting to understand in detail when a similar analysis applies for compact bulk subregions.

In the path integral, the superselection sectors of Sec.~\ref{sec:reducedstate} have a simple origin. The replicas are glued cyclically along the copies of $\phi(a)$, but they all share the same embedded corner $\phi(\partial a)$. So while we integrate over $n$ copies of the bulk fields, there is only a single corner for every $n$, and the representation $R$ of $\mathrm{ECS}$ carried by the fields at the corner must agree across all replicas.

In the representation basis, the path integral $\mathcal{Z}(a^n)$ can be written as
\begin{align}
    \mathcal{Z}(a^n) =\int d\mu(R) \int_{M_n} D\psi_{(n)} e^{-S_n[\psi; R]}
\end{align}
where $S_n[\psi; R]$ is the usual bulk action on the manifold $M_n$, but with the constraints that the fields $\psi$ at $\partial a$ must transform in the $R$ representation of $\mathrm{ECS}$ under a diffeomorphism with support on the corner. The edge mode degrees of freedom at the corner cancel these transformations sector by sector, leaving the action invariant.
This is the representation-theoretic manifestation of our shrinkable boundary condition \eqref{eq:shrinkable}: it ensures that both before and after factorizing the Hilbert space, a gauge transformation at the corner does not affect any correlation function. This can be thought of as a generalization of the constrained path integrals over fixed area states in \cite{Dong:2018seb} to the whole $\mathrm{ECS}$ group.

Crucially, $S_n[\psi;R]$ does \emph{not} include any contributions from sources localized at the corner. Indeed, the boundary conditions near the corner are imposed from a $\epsilon\to 0$ limit of neighborhoods $N_\epsilon$ of the corner, so $\phi(\partial a)$ is never in the integration domain of the path integral. This will be an important point below.

Next, we evaluate the $\psi$ path integral at its dominant saddlepoint, holding the boundary conditions labeled by $R$ fixed. Define $I_n[R]$ to be the on-shell action of this $\psi$-saddlepoint. Then
\begin{align}
    \mathcal{Z}(a^n) \approx \int d\mu(R) e^{- I_n[R]} \,.
\end{align}
To be a fixed-$R$ saddlepoint, the bulk fields $\psi$ must satisfy the equations of motion everywhere, except possibly at the corner $\partial a$.
Focusing on $n=1$ momentarily, and letting $\braket{0} = e^{-I_1}$ be the usual on-shell action for the global state, we can see that
\begin{align}
    e^{-I_1} = \braket{0} = \mathcal{Z}(a) = \int d\mu(R) e^{-I_1[R]} \,. \label{eq:I1def}
\end{align}
Therefore, defining
\begin{align}
    q_R = e^{- I_1[R] + I_1} \,,
\end{align}
we can interpret $q_R$ as a probability distribution over the superselection sectors, normalized with respect to $d\mu(R)$. This is the same distribution $q_R$ appearing in the reduced state \eqref{eq:reducedstate}.
Returning now to the case of general $n$, we can add and subtract $n I_1[R]$ from the action for each fixed $R$. This is useful because the $-n I_1[R]$ cancels $I_n[R]$ in the $n\to 1$ limit. Thus, we can see that
\begin{align}
    \frac{\mathcal{Z}(a^n)}{\mathcal{Z}(a)^n} = \int d\mu(R) e^{- I_n[R] + n I_1[R] - n I_1[R] + n I_1} = \int d\mu(R) q_R^n e^{- I_n[R] + n I_1[R]}\,. \label{eq:ZnoverZton}
\end{align}
Note that the definition of $q_R$ depends on the choice of measure: taking a convention where $I_1$ is fixed (as it should be, because this quantity can be calculated independently of the edge modes), \eqref{eq:I1def} implies that a shift of the measure $d\mu(R) \to d\mu(R) e^{-C_R}$ shifts $q_R \to e^{C_R} q_R$. Thus, the measure dependence of the entropy is still hidden within the distribution $q_R$, just as in the canonical analysis. The Plancherel measure $d\mu(R)$ appears here because we defined the path integral using the flat measure $D\phi$ over embedding fields (as we would usually do other quantum field), to which the Plancherel measure is Fourier dual.

To isolate the area term within the path integral, we split the Plancherel measure as $d\mu(R) = dR\, \mu(R) \equiv dR \, e^{A_R}$, where $dR$ is the microcanonical measure of Sec.~\ref{sec:reducedstate}. Comparing with \eqref{eq:areaop}, $A_R = \ln(\mu(R))$ is precisely the eigenvalue of the area operator $\hat{A}$ in the sector $R$. This is the same split we used to extract the area term in the canonical analysis, adapted to the path integral language. It is also convenient for the saddlepoint approximation, because $dR$ weights the representations uniformly. This is similar in spirit to random matrix theory \cite{Saad:2022rwo}, where the measure over eigenvalues descending from the physical, local measure $dH$ is split into a uniform part $d\lambda$ and a Vandermonde determinant $\Delta(\lambda)$, which is then absorbed into the effective action.

Define $p_R = e^{A_R} q_R$, which is the same distribution over sectors, now normalized with respect to $dR$ instead of the Plancherel measure (this is the distribution $p_R$ of Sec.~\ref{sec:entropy}). Substituting $d\mu(R) = dR\, e^{A_R}$ and $q_R^n = e^{-n A_R} p_R^n$ into \eqref{eq:ZnoverZton}, we find
\begin{align}
    \frac{\mathcal{Z}(a^n)}{\mathcal{Z}(a)^n} = \int dR \, \, p_R^n \, e^{-(n-1) A_R - I_n[R] + n I_1[R]}
\end{align}
so the effective action of the R\'enyi entropy includes a term $(n-1)A_R$ from the measure, which can be thought of as being sourced locally at the corner (because that is where the representation theory data lives). This is why a saddle for a fixed representation $R$ need not be smooth near the corner: the Plancherel measure contributes to the effective action, and can act as a source which can backreact onto the geometry. Note that for fixed area states \cite{Dong:2018seb}, this is exactly the $n$ dependence of $\frac{\mathrm{Area}}{4 G_N}$ in the action.

We now take a saddlepoint approximation in $R$ as well, and assume that the integral localizes onto a particular representation $R_*$. Then the R\'enyi entropy
\begin{align}
    S_n(\rho_a) = \frac{1}{1-n} \ln\left(\frac{\mathcal{Z}(a^n)}{\mathcal{Z}(a)^n}\right)
\end{align}
is well-approximated as
\begin{align}
    S_n(\rho_a) = \frac{I_n[R_*] - n I_1[R_*]}{n-1} + A_{R_*} \,.
\end{align}
Note that the dominant representation $R_*$ will generically depend on $n$. 
Here, we have dropped the R\'enyi analog of the entropy $S(\mathcal{F}(\rho_a))$ of the center which we discussed in Sec.~\ref{sec:centerentropy}. As discussed in Sec.~\ref{sec:centerentropy}, this term is sensitive to the non-perturbative discreteness of the representations. For a distribution $p_R$ whose width scales as a power of $G_N$, it contributes only logarithmically in $G_N$ \cite{Dong:2023xxe}, and we assume it is subleading to the terms we keep. 

Next, we assume that the dominant saddle with action $I_n[R_*]$ satisfies \emph{replica symmetry}: we assume it admits a $\Z_n$ action which cyclically permutes the boundary conditions under which it is invariant. This is not uncommon in AdS/CFT \cite{Lewkowycz_2013}, but see \cite{Akers:2020pmf} for details about what happens when this approximation breaks down. For now, we simply restrict to states where this holds for simplicity, and leave a more detailed analysis for the future. We can now define the orbifolded action
\begin{align}
    I_n[R] = n \widetilde{I}_n[R]  \,, \label{eq:orbifold}
\end{align}
which is computed by the path integral on the $\Z_n$ quotient of the replica symmetric saddle, effectively an $n=1$ computation with boundary conditions at the corner so that \eqref{eq:orbifold} holds.

We can then see that 
\begin{align}
    S_n(\rho_a) = A_{R_*} + \frac{n}{n-1} [\widetilde{I}_n[R_*] - I_1[R_*] ]   \,.
\end{align}
However, note that the boundary conditions defining $\widetilde{I}_n[R_*]$ and $I_1[R_*]$ are identical: both fix the same representation $R_*$ at the corner, and have identical boundary conditions at any other components. So at leading order in $G_N$, the two on-shell actions agree, just as they do for the fixed area states of \cite{Dong:2018seb}. The difference from \cite{Dong:2018seb} is where the area term appears. There, the orbifold action $\widetilde{I}_n[R_*]$ includes the area of the brane imposing the fixed area condition. Here, the shrinkable boundary conditions exclude any such local contribution to the action, and the area instead enters through the measure factor $A_{R_*}$. These two conventions agree if we absorb the measure factor $A_{R_*}$ into the definition of the effective action $I_n[R]$. In our conventions (set by the shrinkable boundary conditions imposed at the corner), the two on-shell actions therefore cancel to leading order, and we find
\begin{align}
    S_n(\rho_a) = A_{R_*}\,.
\end{align}
Comparing with \eqref{eq:entropyformula}, the saddlepoint reproduces the area term $\langle \hat{A} \rangle_\rho = A_{R_*}$ at leading order in $G_N$. The bulk entropy $S_{\mathrm{bulk}}(\rho_a)$ arises at the next order, from fluctuations of $\psi$ around the saddle, and the entropy of the center is the term we dropped above. Similar connections between the area term of the generalized entropy and representation theoretic data associated with the corner of a region have been previously proposed  \cite{Lin:2017uzr,Lin2018JTEntanglement,Blommaert:2018iqz,Mertens:2022ujr,McGough:2013gka,Harlow:2016vwg,Akers:2023fqr,Akers:2018fow,Dong:2018seb,Camps2019Superselection,Lin2026Virasoro,Donnelly:2020teo,Jiang:2020cqo,Balasubramanian:2026ymu,Balasubramanian:2026vlr}. The novelty of this work is to propose a possible direction for deriving this correspondence for more general theories of gravity, and more general subregions, which may not be boundary anchored. 
Note that the same measure factor $A_R$ appears for boundary and bulk subregions. Because the generalized entropy measures the area of the extremal surface for boundary anchored regions, it is plausible that $A_R$ really is the area term of the generalized entropy, at least to leading order. 

\section{The path integral: edge--then--bulk} \label{sec:path_int_2}

In Sec.~\ref{sec:path_int}, we evaluated the path integral computing $\Tr_a[\rho_a^n]$ by first integrating out the bulk fields in each superselection sector $R$, and then evaluating the remaining integral over representations (or, by Fourier duality, the embedding fields) by saddlepoint. In this section, we will instead integrate out the edge modes first, and then examine the resulting path integral over the bulk fields. Assuming  that these two integrals commute, this second procedure  will provide insight into the structure of the replicated geometry which dominates the path integral in the saddlepoint approximation.

When computing the path integral in this order, some spurious infinities appear in the most direct versions of the calculation. These infinities must be regulated and renormalized away, similar to our use of the co-invariant inner product above. In this section, we will instead take a physicist's approach by directly regulating the infinities at intermediate steps of the calculation. As we will see, the final formulas we obtain will be independent of the regulator, which suggests there is a more direct argument to obtain the same results.  We leave the rigorous determination of this more direct argument to future work.

\subsection{Tracing out the edge modes}\label{sec:traceoutedge}

When we cut open $\mathcal{Z}(a)$ or $\mathcal{Z}(a^n)$, we  get a state in the extended Hilbert space which contains edge modes. If we  integrate out the edge modes first, and only then cut open the path integral, we obtain a density operator on the bulk fields alone. In the language of density matrices, integrating out the edge modes is a partial trace over the edge mode factor of $\widetilde{\Ha}_a$. The trace of the resulting operator is then computed by a path integral involving only the bulk fields.

Before performing this partial trace, we need to discuss a subtlety related to the correlations between the edge modes and the bulk fields. After regulating the spacing between representations $R$ to have ``lattice separation'' $e^{- S_0}$,\footnote{The logic of this regulation procedure is as follows. Noncompact groups like ones that appear here have a continuum of representations $R$, and as we discussed in Sec.~\ref{sec:edge} we have to integrate over these.  The divergences in intermediate steps that we mentioned in the preamble to this section arise from the continuum of representations.  So we  regulate the divergence by turning the integral over $R$ into a sum over a lattice of representations spaced by $e^{-S_0}$.  The regulator drops out in the final answer, allowing us to receover the continuum limit.
We denote the regulated lattice spacing as $e^{-S_0}$ because the state-independent constant ambiguity in \eqref{eq:entropyformula} arises precisely because of the scheme dependence sending $S_0 \to S_0 + \lambda$, so we can think of $S_0$ as a counterterm for the entropy of a subregion.} the projectors $\Pi_R$ satisfy
\begin{align}
    \Pi_R \Pi_T = \frac{e^{S_0}\delta_{R T}}{\mu(R)} \Pi_R\,,\label{eq:Pi_R_regulated}
\end{align}
where $\delta_{R T}$ is the Kronecker delta function between representations, $e^{S_0} \delta_{R T}$ is the regulated delta function over representations, and $\mu(R)$ is the Plancherel weight \eqref{eq:muRdef} of each representation. Note that when $G$ is compact, $\mu(R)$ is the dimension of $V_R$, the vector space of the $R$ representation, and $e^{S_0}$ is the volume of $G$. 

The reduced state $\rho_a$ can then be written as
\begin{align}
    \rho_a = \sum_R e^{-S_0} p_R \rho_R \otimes \Pi_R \,, \qquad \rho_a^n = \sum_R e^{-S_0} \mu(R)^{1-n}\, p_R^n\, \rho_R^n \otimes \Pi_R\label{eq:rhoan_compact}\,.
\end{align}
Here, $p_R$ is a probability distribution such that $\sum_R e^{-S_0} p_R = 1$, so that in the $S_0\to\infty$ limit, we have that $\int dR \, p_R = 1$. 
In order to take the partial trace over the edge modes, we need to lift the constraint that the representation $R$ between the edge modes and the bulk fields must agree. In other words, we need to define an isometric map
\begin{align}
    \mathcal{V}: \widetilde{\Ha}_a \into \Ha_{\mathrm{tot}} =  \bigoplus_{R,T} \Ha_a(R) \otimes V_T\,,
\end{align}
where $V_T$ is the vector space of the $T$ representation, and the extended Hilbert space $\widetilde{\Ha}_a$ embeds as the diagonal subspace where $R = T$. This is necessary because we defined the extended Hilbert space $\widetilde{\Ha}_a$ as the direct sum over $\Ha_a(R) \otimes V_R$, which does not tensor factorize between the bulk fields and the edge modes. 
Actually, to renormalize the $S_0$ divergence in the $S_0\to \infty$ limit where we remove our regulator, we should define $\mathcal{V}^\dagger\mathcal{V} = e^{S_0} \Id$, so $\mathcal{V}$ is only projectively an isometry. This has the effect of  renormalizing the entropy of $\mathcal{V}\rho_a\mathcal{V}^\dagger$ by the constant $S_0$, so the $S_0\to\infty$ limit is finite.

More precisely, we define the trace on $\Ha_{\mathrm{tot}}$ to be
\begin{align}
    \Tr_{\mathrm{tot}}[\mathcal{O}] = \sum_{R,T} e^{- S_0} \Tr_{\Ha_a(R) \otimes V_T}[\mathcal{O}]\,,
\end{align}
which is the usual definition of the trace on $\Ha_{\mathrm{tot}}$, renormalized by $e^{- S_0}$. Then since $\mathcal{V}$ is a (projective) isometry,  $\Tr_a[\rho_a] = \Tr_{\mathrm{tot}}[\mathcal{V}\rho_a\mathcal{V}^\dagger]$, and that $(\mathcal{V} \rho_a \mathcal{V}^\dagger)^n = e^{(n-1) S_0}\mathcal{V} \rho_a^n \mathcal{V}^\dagger$. In particular, this implies that $\rho_a$ and $\mathcal{V} \rho_a \mathcal{V}^\dagger$ have the same R\'enyi entropy for any $n$, up to the constant $S_0$. Essentially, similar to the factorization map $V$ we used in Sec.~\ref{sec:reducedstate}, $\mathcal{V} \rho_a \mathcal{V}^\dagger$ embeds $\rho_a$ into a factorizing Hilbert space $\Ha_{\mathrm{tot}}$. The difference is that while the edge modes introduced within the $R$ sector account for the degeneracy \emph{within} each representation, which led to an additional contribution to the entropy, because each representation $R$ appears uniquely in the sum, there is no degeneracy \emph{between} the representations. Thus, the ``edge modes'' introduced by $\mathcal{V}$ are trivial. 

We can then define the coarse-graining map which traces out the edge modes as
\begin{align}
    \overline{\mathcal{O}} := \Tr_{\mathrm{edge}}[\mathcal{V}\mathcal{O}\mathcal{V}^\dagger] \,.
\end{align}
To match \eqref{eq:physicaltrace}, we must continue to formally define $\tr[\Pi_R] = 1$ within each representation. Note that this coarse-graining map is linear and trace preserving.
In particular, the reduced state is mapped to
\begin{align}
    \overline{\rho_a} = \bigoplus_R p_R \rho_R \,, \qquad \overline{\rho_a^n} = \bigoplus_R \mu(R)^{1-n}\, p_R^n\, \rho_R^n  \label{eq:coarsegrainedstates}\,.
\end{align}
These definitions contain no factors of $S_0$ because $\mathcal{V}$ is a projective isometry, as opposed to an isometry.
Each coarse-grained state $\overline{\rho_a}$ is defined on the bulk fields alone, and is the state we would get by  integrating out the edge modes before cutting open the path integral. Note in particular, however, that $\overline{\rho_a^n} \neq (\overline{\rho_a})^n$. More precisely, we can see that if we define an operator
\begin{align}
    \hat{A} = \bigoplus_R \ln(\mu(R)) \Id_{\Ha_a(R)} \,,
\end{align}
which is essentially the same area operator as in \eqref{eq:areaop} but restricted to the bulk fields, we have that 
\begin{align}
    \overline{\rho_a^n} = e^{-(n-1)\hat{A}}(\overline{\rho_a})^n = e^{\hat{A}} (e^{-\hat{A}} \,\overline{\rho_a})^n \,.
\end{align}
Thus, it will be useful to define the density matrices
\begin{align}
    \rho_{\mathrm{can}} &= \overline{\rho_a}\,,\\
    \rho_{\mathrm{gen}} &= e^{-\hat{A}} \rho_{\mathrm{can}} \,.
\end{align}
Following the terminology of \cite{Akers:2023fqr}, we refer to $\rho_{\mathrm{can}}$ as the \emph{canonical} density matrix, and $\rho_{\mathrm{gen}} $ as the \emph{generalized} density matrix. Similarly, we define the canonical and generalized trace as $\Tr_{\mathrm{can}}[\,\cdot\,] = \sum_R e^{-S_0}\tr_{\Ha_a(R)}[\,\cdot\,]$, and $\Tr_{\mathrm{gen}}[\,\cdot\,] = \Tr_{\mathrm{can}}[e^{\hat{A}}\,\cdot\,]$, respectively.\footnote{Note that in the $S_0 \to \infty$ limit, we have that $\sum_R e^{- S_0} \to \int dR$, so this definition of canonical trace has a well-defined limit in the continuum limit.} These are the traces such that
\begin{align}
    \Tr_a[\rho_a] = \Tr_{\mathrm{can}}[\rho_{\mathrm{can}}] = \Tr_{\mathrm{gen}}[ \rho_{\mathrm{gen}}] = \sum_R e^{- S_0} p_R \tr_{\Ha_a(R)}[\rho_R] \,. \label{eq:manytraces}
\end{align}
More generally, it is easy to show from the definitions that
\begin{align}
    \Tr_a[\rho_a^n] = \Tr_{\mathrm{can}}[\overline{\rho_{a}^n}] =\Tr_{\mathrm{can}}[e^{-(n-1)\hat{A}}\,(\rho_{\mathrm{can}})^n] = \Tr_{\mathrm{gen}}[ \rho_{\mathrm{gen}}^n]\,.
\end{align}

As a reminder, $\rho_a$ is the state on the extended Hilbert space for $a$, which includes the edge modes. $\rho_{\mathrm{can}}$ and $\rho_{\mathrm{gen}}$ are states which are defined purely from the bulk fields. $\rho_{\mathrm{can}}$ is the state we obtain by cutting open the path integral in the usual way (because $\Tr_{\mathrm{can}}$ sums uniformly over all the sectors $R$ which appear in the Hilbert space). However, as we can see, it is the generalized trace $\Tr_{\mathrm{gen}}$, and its associated state $\rho_{\mathrm{gen}}$, which correctly reproduces the entropy of the full state $\rho_a$. We can use these traces to explicitly compute the $n$th R\'enyi entropy of these density matrices, and take the $n\to1$ limit to compute the von Neumann entropy. We see that
\begin{align}
    S(\rho_a) = S(\rho_{\mathrm{gen}}) = \langle \hat{A} \rangle + S(\rho_{\mathrm{can}}) \,. \label{eq:S_coarsegrained}
\end{align}
Thus, one way to think about the area term is that it is the entropy which arises from coarse-graining of the edge modes on the corner.

We can also phrase the relationship between these density matrices in terms of their modular Hamiltonians. Let $K_{\mathrm{gen}} := -\ln \rho_{\mathrm{gen}}$ and $K_{\mathrm{bulk}} := -\ln \rho_{\mathrm{can}}$. By \eqref{eq:rhoan_compact}, within each sector $R$ we have $K_{\mathrm{gen}} = \big(\ln \mu(R) - \ln(p_R\rho_R)\big)$. Thus, we can see that
\begin{align}
	K_{\mathrm{gen}} = \hat{A} + K_{\mathrm{bulk}}\,, \label{eq:JLMS}
\end{align}
Eq.~\eqref{eq:JLMS} is precisely the relationship between the boundary and bulk modular Hamiltonians in AdS/CFT \cite{Jafferis_2016}, with $\hat{A}$ playing the role of the area operator, and taking its expectation value in $\rho_a$ reproduces \eqref{eq:S_coarsegrained}. Note that $K_{\mathrm{bulk}}$ includes the term $-\ln p_R$ from the center, just as in the algebraic formulation of \cite{Harlow:2016vwg}. This suggests that the extended Hilbert space of the subregion acts as an effective boundary, as in AdS/CFT. For this identification to make sense, $S(\rho_a)$ must be bounded below, which requires the spectrum of representations to be discrete \cite{Balasubramanian:2026ymu}. 

Finally, note that we have only traced out the $\mathrm{ECS}$ edge modes which arise from locally factorizing the perturbative Hilbert space. Non-perturbative effects, such as a wormhole connecting $a$ to its complement, would further obstruct factorization, and could lead to a more intricate edge mode structure \cite{Klinger:2026dhz}.

\subsection{The reduced state in the group basis}

To gain more intuition for the coarse-graining of Sec.~\ref{sec:traceoutedge}, we now rewrite $\rho_a$ and its powers as sums over the group elements $g$ (corresponding to embeddings of the corner) rather than by representations of the gauge group. To begin, consider the projector $\Pi_R$ onto the $R$ representation in the edge mode sector. Define $d_R = e^{-S_0} \mu(R)$, which is the relative weight of $\Pi_R$ with respect to the counting measure over representations. In the representation basis of the Peter--Weyl (or Plancherel) theorem, this projector decomposes as
\begin{align}
    \Pi_R = \frac{1}{d_R}\sum_{i} \ketbra{R,i} \,.
\end{align}
The states inside this sum are normalized such that 
\begin{align}
    \braket{R,i}{R',j} = \delta_{ij} \delta_{R,R'}\,, \label{eq:planchnorm}
\end{align}
where $\delta_{R,R'}$ is the Kronecker delta function. We can also define the operator 
\begin{align}
    \widetilde{\Pi}_R = \int dg \,\chi_R(g^{-1}) u(g) \,,\label{eq:PiRtilde}
\end{align}
where $u(g)$ is the unitary representation of the gauge group action in each sector $R$ of the edge mode Hilbert space. More precisely, if $u_R(g)$ is the unitary matrix for the $R$ representation, then
\begin{align}
    u(g) = \sum_R \,u_R(g) \,. \label{eq:udecomp}
\end{align}
In fact, $\Pi_R$ and $\widetilde{\Pi}_R$ are the same operator. To see this, note that we can expand $\widetilde{\Pi}_R$ as
\begin{align}
    \widetilde{\Pi}_R =\int dg \sum_{R'} \,\chi_R(g^{-1}) \, u_{R'}(g) \,.
\end{align}
Expanding $u_{R'}(g)$ in a complete set of states
\begin{align}
    \widetilde{\Pi}_R =\sum_{ij}\int dg \sum_{R'} \,\chi_R(g^{-1})\, u_{R'}(g)_{ij} \ketbra{R',i}{R',j} \,.
\end{align}
We then do the $g$ integral to see\footnote{In this expression, we used the fact that $\braket{\chi_R}{R',ij}_{L^2(G)} = \delta_{ij} \delta(R,R')$ in $L^2(G)$, and $\delta(R,R') = e^{S_0} \delta_{R,R'}/\mu(R) = \delta_{R,R'}/d_R$ with our regulator. }
\begin{align}
    \widetilde{\Pi}_R &= \sum_{ij}\sum_{R'}\,\braket{\chi_R}{R',ij}_{L^2(G)}  \,\ketbra{R',i}{R',j} \,,
    \\&= \sum_{i} \sum_{R'}\,\frac{\delta_{R,R'}}{d_R} \ketbra{R',i}{R',i}
    \\&= \Pi_R \,.
\end{align}
We can use this group-basis presentation of $\Pi_R$ to rewrite the state $\rho_a$.
We define the fixed-group element density matrix as the smearing of $\rho_R$ with respect to the character function: 
\begin{align}
    \rho[g] := \bigoplus_R \,\mu(R) \chi_R(g^{-1})\, q_R \rho_R  \,. \label{eq:rhog}
\end{align}
Recall that $q_R = p_R / \mu(R)$, as in \eqref{eq:reducedstate}. This normalization will be useful below.

With this definition, and substituting \eqref{eq:PiRtilde} into \eqref{eq:rho_a}, we see that
\begin{align}
    \rho_a = \int dg \, \rho[g] \otimes u(g) \,. \label{eq:rhoa_gbasis}
\end{align}
At first, this expression looks quite analogous to \eqref{eq:reducedstate}. However, there are some key differences. First, using the fact that $\rho_R$ is Hermitian and $\chi_R(g)^* = \chi_R(g^{-1})$, we can see that
\begin{align}
    \rho[g]^\dagger = \rho[g^{-1}]\,.
\end{align}
This means that unless $g = g^{-1}$, $\rho[g]$ is not Hermitian. This is actually a good thing, because $u(g)$ is not Hermitian either. Indeed, the reason $\rho_a$ is Hermitian is because
\begin{align}
    \rho_a^\dagger = \int dg \,\rho[g^{-1}] \otimes u(g^{-1}) = \rho_a\,,
\end{align}
where we used the invariance of the Haar measure under inversions.

Let us discuss some more properties of $\rho[g]$, before proposing a path integral interpretation of this object. First, note that because the character is a class function (constant on conjugacy classes, by cyclicity of the trace), $\rho[g]$ is as well:
\begin{align}
    \rho[g] = \rho[hgh^{-1}] \,. \label{eq:defectclassfn}
\end{align}
This will be important in the path integral interpretation of $\rho[g]$ below.

To interpret $\rho[g]$, we begin by discussing the fixed area state analysis of \cite{Dong:2018seb}, where we neglect the normal translations which can move the corner around, and only quantize the edge mode associated with uniform boosts around the $\mathrm{HRT}$ surface. In that case, one can prepare a fixed area state (which we will refer to as having an area $A_R$) by inserting a delta function
\begin{align}
    \tr[\rho_R] = Z[A_R] = \int D\psi d\lambda e^{-I_1[\psi]}\, e^{i\lambda (A(\psi)-A_R)} \,.
\end{align}
Here, $A(\psi)$ is the area of the corner (which we have fixed to be an extremal surface), and $I_1[\psi]$ is the usual Einstein-Hilbert action, plus matter fields. The full path integral is then computed as
\begin{align}
    \tr[\rho_a] = \int dA_R \,Z[A_R] \,.
\end{align}
Then, using \eqref{eq:rhog}, and that the character of the group of boosts around the corner (isomorphic to $\R$) is just $\chi_R(\theta) = \exp(i A_R \theta)$, where $\theta$ is the boost angle around the corner which defines the group element, the trace of $\rho[\theta]$ (the analog of $\rho[g]$ in this setting) is given by
\begin{align}
    \tr[\rho[\theta]] = \int dA_R  \,e^{i A_R \theta} \int D\psi d\lambda\, e^{-I_1[\psi]} e^{i\lambda (A(\psi)-A_R)} = \int D\psi \,e^{-I_1[\psi]}  e^{i\theta A(\psi) }\,.
\end{align}
This can be interpreted as the effective action of Einstein gravity, along with a brane of tension $i\theta$ which has been placed at the extremal surface. If we solve the Einstein equations with this source,  backreaction causes a conical defect to appear, localized at the location of the brane \cite{Dong:2018seb}. The strength of this conical defect is proportional to the tension $\theta$. If we slice open this path integral on a time symmetric slice which ends on the extremal surface to prepare $\rho[\theta]$, \eqref{eq:rhog} has the interpretation of a state with a relative conical defect $\theta$ between the states on the ``bra'' and the ``ket'' of $\rho[\theta]$.

Returning to the case of the full corner symmetry group, recall that the group element $g$ represents an equivalence class of diffeomorphisms which can possibly move the corner within the spacetime. We propose that $\rho[g]$, computed as in \eqref{eq:rhog}, represents a spacetime with a defect $g$ at the corner $\partial a$, analogous to the fixed area state example above. The defect is such that there is a relative shift by the diffeomorphism $g$ between the fields on the ``bra'' and the fields on the ``ket'' of $\rho[g]$. Because $g \neq g^{-1}$ in general, these boundary conditions are not $\mathrm{CRT}$ symmetric, and therefore $\rho[g]$ will not be Hermitian.

For this interpretation to be consistent, the monodromy $g$ should be invariant under simultaneous $\mathrm{ECS}$ transformations of the bra and the ket. This is because we could have defined the subregion $A = \phi(a)$ using any initial coordinate system we like. If we transformed the bra and ket simultaneously by a corner symmetry transformation $h$, then we could equivalently start with the initial configuration and implement a monodromy $h^{-1} g h$ as we circle the corner in the $t-r$ plane. However, recall that because of \eqref{eq:defectclassfn}, $\rho[g]$ is precisely invariant under this transformation. So we should think of $[g]$ as a \emph{relative} $\mathrm{ECS}$ transformation between the bra and the ket preparing $\rho[g]$. One can trace this result as a consequence of the fact that simultaneous corner symmetry transformations leave the global state $V\ket{0}$ invariant.
Because of this, $[g]$ is a physically well-defined quantity, which is a nice consistency check.

We can extend this interpretation to $\rho_a^n$ as follows. Using $u(g)u(h) = u(gh)$ in \eqref{eq:rhoan_gbasis}, we can write\footnote{Technically, this group-like property of $u(g)$ should only in the limit $S_0\to\infty$. However, note \eqref{eq:rhong_character} provides an independent definition of $\rho_n[g]$ which does not use this group-like property, and agrees with this naive definition after regulating appropriately, so \eqref{eq:rhoan_gbasis} is still well-defined in the regulated theory.}
\begin{align}
	\rho_a^n = \int dg\, \rho_n[g] \otimes u(g)\,, \label{eq:rhoan_gbasis}
\end{align}
where $\rho_n[g]$ is the $n$-fold convolution of $\rho[g]$ with itself,
\begin{align}
	\rho_n[g] := \int dh_1 \cdots dh_n\, \delta(g^{-1} h_1 \cdots h_n)\, \rho[h_1] \cdots \rho[h_n]\,. \label{eq:rhongconvolution}
\end{align}
Expanding the definition of each $\rho[g]$, this is equivalent to 
\begin{align}
	\rho_n[g] = \bigoplus_R \mu(R)\, \chi_R(g^{-1})\, q_R^n\, \rho_R^n\,, \label{eq:rhong_character}
\end{align}
which reproduces \eqref{eq:rhoan_compact} upon substituting into \eqref{eq:rhoan_gbasis}.
Note that more generally, for any $n,m$, we have that
\begin{align}
	\rho_{n+m}[g] = \int dh \rho_{n}[g h^{-1}] \rho_{m}[h] \,. \label{eq:convolutions}
\end{align}
This has a nice interpretation: $\rho_a^{n+m}$ can be computed by first doing the path integral on the first $m$ replicas, then doing the path integral on the final $n$ replicas, and gluing together the results via the convolution of the relative defects gives the $(n+m)$-fold replica of the path integral. This is just the prescription that we should identify the embeddings on the bra of $\rho_n$ and ket of $\rho_m$ and integrate over the possible embeddings, as with any other quantum field. Again, we expect this structure to hold perturbatively in $G_N$, perhaps unless we add more edge modes for wormholes as in \cite{Klinger:2026dhz}. Equation \eqref{eq:convolutions} also provides another consistency check that $\rho[e]$ is normalized correctly to produce a defect free spacetime: regardless of the replica number $n$, there are no additional defects or $n$-dependence that appears when gluing together the partial replicas and setting $g=e$.

Combining \eqref{eq:rhong_character} with \eqref{eq:manytraces}, we find
\begin{align}
	\Tr_a[\rho_a^n] = \Tr_{\mathrm{gen}}[\rho_{\mathrm{gen}}^n] = e^{-S_0}\Tr_{\mathrm{can}}[e^{- \hat{A}}\, \rho_n[e]] \,. \label{eq:renyi_defectfree}
\end{align}
So, after renormalizing away the state independent constant $S_0$ (which arises because the factor of $\chi_R(e) = e^{S_0} \mu(R)$ within the definition of $\rho_n[e]$), and weighting the trace by $e^{-\hat{A}}$, the $n$th R\'enyi entropy only recieves contributions from replicated spacetimes which are defect-free.

This factor of $e^{-S_0}$ arises because the volume of each conjugacy class of a non-compact group is divergent. So one way to implement this renormalization scheme is to define the (complex valued) measure $d\nu([g]) = \tr[u(g)] dg$, in which case we can define the renormalized ``convolution trace'' $\Tr_{\mathrm{conv}}$ as
\begin{align}
    \Tr_a[\rho_a^n] = \Tr_{\mathrm{conv}}[\rho_n[e]] := \int d\nu([g]) \Tr_{\mathrm{can}}[\rho_n[g]] \,.
\end{align}
Here, we take $d\nu([g])$ to have support over the \emph{conjugacy classes} of $G$, because this is the group-theoretic data that $\rho[g]$ depends on. In other words, $d\nu([g])$ simply restricts the domain of integration so that the $e^{S_0}$ redundancy $[g] \to [hgh^{-1}]$ does not arise when taking the trace, which is why this is the renormalized definition. But at least formally, this is the same definition as \eqref{eq:renyi_defectfree}. It would be interesting to work this out in more detail, but for now we will simply take the definition of $\rho_n[e]$ as suggestive interpretation for $\Tr_{\mathrm{gen}}[\rho_{\mathrm{gen}}^n]$, which is well-defined for non-compact groups.

\subsection{Which subregion?} \label{sec:subregion}

In AdS/CFT, the entanglement wedge $E(A)$ of a boundary subregion $A$ can be defined as the bulk subregion which is accessible to boundary operators within $A$ under repeated and smeared modular flow \cite{Faulkner:2017vdd,Cotler:2017erl}. In other words, with $K_A = - \ln(\rho_A)$,  boundary operators of the form $\mathcal{O}(t) = e^{i K_A t} \mathcal{O}(0) e^{-i K_A t} $  for any $t$ will be dual to an operator with support in $E(A)$. Conversely, any bulk operator within $E(A)$ can be reconstructed (at least within the code subspace that it is defined for) by smearing the modular flowed operator $\mathcal{O}(t)$ against some (generally unknown) kernel, which is essentially a generalization of the HKLL procedure \cite{Hamilton:2006az}. A generalized entropy is then associated with the entanglement wedge $E(A)$.

We would like to know if we can define a bulk region $E(a)$ which has this property.\footnote{We will abuse notation and use  $a,E(a)$ to refer to a spatial subregion and its Lorentzian domain of dependence interchangeably, because operators in the domain of dependence are related to operators on the spatial slice by unitary time evolution.} In other words, for an operator $\mathcal{O}$ with support on $a$, we would like to know what is the spacetime region accessible to the (smearing and completion of) all operators of the form $\mathcal{O}(t) = \rho_a^{it} \mathcal{O} \rho_a^{-it}$.\footnote{For quantum fields, this definition can suffer from UV divergences, and needs to be canceled by the inverse modular flow with respect to $\rho_{\mathcal{a}}$. We will neglect this subtlety, and a more careful treatment of these issues in the future would be interesting.} 

Note that $\rho_a$ includes a contribution from the edge modes, so the edge modes will affect the modular flow of the operators in $a$. In this paper, the edge modes introduced in the extended Hilbert space have played several roles. First, they allowed us to define the finite region $\phi(a)$ in a diffeomorphism invariant way, through the embedding field $\phi$. They also contribute directly to the entropy \eqref{eq:entropyformula} through the expectation value of the area operator $\hat{A}$. In the previous subsection, they also have another, more speculative role: integrating over the edge modes allows us to glue together partial replicas $\rho[g]$ to compute the $n$-fold replica $\rho_n[g]$, which can be thought of as arising from integrating over the embedding of $a$ between the bra and ket of sequential replicas.

Because of this final role of the edge modes, it is worth asking whether the modular flow under $\rho_a$ can cause the region to ``move'' in a spacelike direction. Indeed, the corner symmetry group $\mathrm{ECS}$ includes the $\R^2$ normal translations which move the corner within the spacetime, and so the relative defects $h_i$ carried by the partial replicas in \eqref{eq:rhongconvolution} are not restricted to transformations which fix the corner. It is tempting to speculate that in the dominant saddle, the partial replicas may include such a translation, so that the fixed locus of the replica symmetry lies not at $\partial a$ but at the boundary of a different region. 

To support this intuition, suppose that $\rho_n[e]$ is dominated by partial saddles $\rho[h_1], \cdots ,\rho[h_n]$.  Define $g_1 = h_1$, $g_2 = h_2 h_1$, ..., $g_n = h_n \cdots h_1 = e$. Then as we discretely twist the dominant saddle around the replicated direction, an operator $\mathcal{O}$ with support in $a$ will now have support within $g_1 \cdot a$, $g_2 \cdot a$, ... $g_n \cdot a = a$. This is because each $\rho[h]$ describes a partial saddle with a relative defect $[h]$ between its bra and ket. In general, even if the original operator $\mathcal{O}$ was local, this flow can produce a highly non-local operator. This ``discretely translated'' operator is the Euclidean version of the modular flowed operator.

So, we define $\widetilde{E}_n(a) = \cup_{i=1}^n  \phi_{g_i}(a)$. Here, by $\phi_e(a)$ we mean the original subregion that we used to define the quantum state $\rho_a$ in the first place, and $\phi_g(a)$ is $\mathrm{ECS}$ transform of $\phi_e(a)$ by $g$. We will abuse notation and write $a$ instead of $\phi_e(a)$ from now on. Assuming we can analytically continue the corresponding subregion in $n$, we then define $\widetilde{E}(a) = \lim_{n\to1} \widetilde{E}_n(a)$. 

$\widetilde{E}_n(a)$ is the union of all the ``replicated spatial slices'' that an operator from the original $n=1$ spacetime can see. In other words, although $a$ is gauge invariantly defined, because simultaneous diffeomorphisms between the bulk fields and the edge modes cancel, there can be fluctuations of the embedding field which effectively enlarge the region as we partially twist around the replica. Equivalently, we can measure fluctuations relative to the embedding field, in which case the fields outside of $a$ can flow into the region $a$ itself, and vice versa. But because $a \subset \widetilde{E}_n(a)$ for all $n$, the union over replicated subregions accessible by these fluctuations will always at least include the original region, even if a particular partial saddle includes an ``inward'' fluctuation which shrinks the region: the constraint that the total replicated spacetime is defect-free forces that the relative defect of another partial saddle must cancel this inward fluctuation. 

Because $\Tr[\rho_a^n] \sim \Tr[e^{- K_a n}]$ and we have just argued that each partial saddle which contributed to $\Tr[\rho_a^n]$ may reach out beyond the original subregion, it is plausible that modular flow using $K_a$ may also allow operators in the $n=1$ spacetime to ``reach out'' beyond the original subregion. Because the area operator $\hat{A}$ is central within the algebra of operators in the subregion, and commutes with $K_a$, it must measure the area of the boundary of the subregion that $a$ can influence even after it may ``reach out'' past its original definition. In other words, the area operator in the entropy formula \eqref{eq:entropyformula} should measure the area of the entanglement wedge $E(a)$. The partial saddle analysis suggests that the bulk subregion accessible to an operator in $a$ will be $\widetilde{E}(a)$. 

Therefore, we propose that $E(a) = \widetilde{E}(a)$. In other words, we propose that the entanglement entropy of $\rho_a$ is computed as
\begin{align}
    S(\rho_a) = \min_{E(a) \supseteq a}\left[\frac{\mathrm{Area}(\partial E(a))}{4 G_N} + S(\mathcal{F}(\rho_a)) + S_{\mathrm{bulk}}(\rho_a)\right]\,. \label{eq:BPentropy}
\end{align}
This matches the generalized entanglement wedge proposal of Bousso and Penington \cite{Bousso:2022hlz,Bousso:2023sya}, which we briefly review here. On a static Cauchy slice, \cite{Bousso:2022hlz} proposed that a finite subregion in gravity can encode the semiclassical physics of a larger bulk subregion $E(a)$, called the generalized entanglement wedge of $a$. $E(a)$ is defined as the region containing $a$ which minimizes the generalized entropy. This is analogous to the relationship between a boundary subregion $R$ and its entanglement wedge $E(R)$ in AdS/CFT. Our proposal is that the edge modes of the region, and in particular the normal translations of the corner, provide a \emph{mechanism} for the enlargement of the generalized entanglement wedge, compared to the original region.

Note that we have not shown that such saddles with a displaced corner exist, or that they are compatible with the boundary conditions defining $\rho_a$. Establishing this would require solving for the replicated geometry with the relative defects $h_i$ described above, confirming that the Euclidean modular flow argument works as expected, and establishing that the dominant saddle indeed satisfies $\widetilde{E}(a) = E(a)$. We leave this construction, which would constitute a rigorous derivation of our intuitive argument above, for future work. 

Finally, we note that the relation $\rho_{\mathrm{gen}} = e^{-\hat{A}}\rho_{\mathrm{can}}$ between the generalized and canonical density matrices of Sec.~\ref{sec:traceoutedge} is the same as the relation proposed for boundary subregions and their entanglement wedges in \cite{Akers:2023fqr}. In our construction, this relation follows from tracing out the edge modes. The generalized density matrix plays an important role in the one-shot analysis of \cite{Akers:2020pmf,Akers:2023fqr}, which refines the standard story of entanglement wedge reconstruction to a more general class of states. It was conjectured in \cite{Bousso:2023sya} that this more general structure is necessary to extend the generalized entanglement wedges of \cite{Bousso:2022hlz} to dynamical spacetimes. Because of the similarities between the density matrices $\rho_{\mathrm{gen}}$ and $\rho_{\mathrm{can}}$ to the one-shot analyses of \cite{Akers:2020pmf,Akers:2023fqr}, the formalism we proposed in this paper seems to be capable of supporting this more general structure as well, suggesting that our analysis may be generalizable to dynamical spacetimes. It would be interesting to make this connection more precise. Because the one-shot structure between $\rho_{\mathrm{gen}}$ and $\rho_{\mathrm{can}}$ follows from the reduced density matrix $\rho_a$, which itself follows from the factorization of the Hilbert space rather than from the saddlepoint analysis, there is some hope that our analysis could be extended to dynamical spacetimes.

\section{Discussion}\label{sec:disc}

In this paper, we investigated the entanglement entropy of a finite region $a$ of a gravitating spacetime at a moment of time symmetry. The central ingredient was an embedding field $\phi: a \into \Sigma$ which specifies the region in question in a diffeomorphism invariant way. The embedding field $\phi$ plays the role of the gravitational edge mode, transforming under the extended corner symmetry group $\mathrm{ECS}$ of diffeomorphisms at the corner $\partial a$. Assuming there is a UV regulator for the diffeomorphism group which renders $\mathrm{ECS}$ locally compact, we showed in Sec.~\ref{sec:edge} that the entropy of the subregion $a$ decomposes into three pieces, each of which is an integral over representations $R$ of the corner symmetry group, in \eqref{eq:entropyformula}. These three pieces are the classical uncertainty of the distribution over representations, the expectation value of the operator $\hat{A}$ which is fixed by the Plancherel measure of $\mathrm{ECS}$, and a bulk entropy term. In 3D gravity in AdS spacetimes, the eigenvalues of $\hat{A}$ reproduce $\ell/4G_N$ \cite{Balasubramanian:2026vlr}. In this paper, we proposed that $\hat{A}$ measures the area of a co-dimension 2 surface more generally. In Sec.~\ref{sec:path_int}, we proposed a gravitational path integral which prepares $\rho_a$ that includes an integral over $\phi$, and showed that its saddlepoint reproduces the area term at leading order in $G_N$. In Sec.~\ref{sec:path_int_2}, we showed that integrating out the edge modes first led to a structure analogous to the generalized density matrix of \cite{Akers:2023fqr}. Finally, in Sec.~\ref{sec:subregion} we argued that when $\partial a$ is not extremal, the dominant replicated saddle can displace the corner, so that the area term measures the boundary of the generalized entanglement wedge $E(a)$ of \cite{Bousso:2022hlz,Bousso:2023sya} rather than $\partial a$ itself. In other words, we argued that \eqref{eq:entropyformula} and \eqref{eq:BPentropy} are equivalent.
The explicit construction of these saddles, however, was left to future work.

Several recent works have obtained results related to ours.  In \cite{Folkestad:2023cze}, it was argued that extremal surfaces and non-perturbative features of the background are sufficient structures to dress perturbations to allow for independence of subregions in quantum gravity. Our findings are consistent with this result. Assuming that the integration over embeddings can indeed move the corner within the dominant saddle prepating the reduced state, there are two cases to consider. When $\partial a \neq \partial E(a)$, $\partial E(a)$ is extremal \cite{Bousso:2022hlz}, and according to \cite{Folkestad:2023cze}, dressing to this portion of the corner is allowed. On the other hand, if $\partial a = \partial E(a)$, then this portion of the corner is fixed to all orders in perturbation theory, i.e., it is non-perturbatively defined, so we can also dress the fields in the subregion to this portion of the corner. 

The authors of \cite{Dong_2020} define a notion of ``effective entropy'' for a subregion in a gravitating spacetime. However, they arrive at a different formula for the entropy, as they only consider area contributions when $\phi(a)$ contains a different number of components than $a$, i.e. only the area of islands. In contrast, we also have an area contribution from the original region $a$ as well.\footnote{More precisely, perhaps, its generalized entanglement wedge $E(a)$.} Additionally, they did not fully specify how to define a subregion in a gauge invariant way. In contrast, we used the embedding map $\phi$ to define the subregion $\phi(a)$ in a manifestly gauge invariant way. It would be interesting to see if their arguments can be adapted to include edge modes of the subregion, which may allow our formalism to capture entanglement islands as well.

Finally, in \cite{Jensen:2023yxy,Chandrasekaran:2022cip,Kudler-Flam:2023qfl} the authors compute the entropy of a subregion in a gravitating spacetime by assuming the existence of an observer and dressing the fluctuations in $a$ to this observer. Using algebraic techniques such as the crossed product, they have all found that the algebra of observables undergoes a phase transition from type III, as is the case with local QFT, to type II, which has a well-defined entropy up to an overall state-independent ambiguity. Note that our entropy formula for $S(\rho_a)$ is also only well-defined up to a state-independent constant. As we have emphasized in the main text, there is a suggestive analogy between the edge modes we consider here and the observers considered in \cite{Jensen:2023yxy,Chandrasekaran:2022cip,Kudler-Flam:2023qfl}. Roughly, the authors of these articles impose the constraint between the bulk fields and the edge modes for a subset of the edge modes which transform in some subalgebra of the $\sl(2,\R)$ component of $\ecs$. On the other hand, our argument for the larger region computing the entropy relied  on including the $\R^2$ component of $\ecs$. This suggests that it is the $\sl(2,\R)$ part of $\ecs$ which makes the entropy UV finite, and the $\R^2$ part which determines its actual value. The explicit construction of an algebraic picture connecting edge modes with the crossed product has been investigated in \cite{Klinger:2023tgi}, and it would be interesting to make the connection of these results to our own more precise in future work.

\subsection{Applications} \label{sec:applications}

\begin{figure}
    \centering
    \includegraphics[]{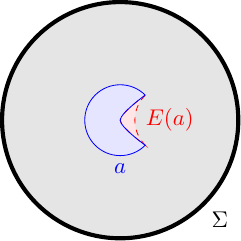}
    \caption{On a static slice $\Sigma$ of AdS, a causally complete region a and its entanglement wedge E(a), which includes the blue region as well. In $2+1$d, the dashed red line is a geodesic. In higher dimensions, it is a minimal surface.}
    \label{fig:bulkew}
\end{figure}

We end with some examples of what the Bousso--Penington proposal implies. Let $\Sigma$ be asymptotically AdS or flat, and let $a$ have a single connected component. Additionally, assume the bulk matter fields are in their ground state and that $\Sigma$ is two dimensional, i.e., is a Cauchy slice for 2+1 dimensional spacetime. We will now show that $E(a)$ is the convex hull of $a$.\footnote{By convex hull, we mean the smallest region $E(a)$ containing the original region $a$ such that any two points in $E(a)$ are connected by a geodesic.} First, note that it is sufficient to prove that any two points on $\partial E(a)$ are connected by a geodesic passing entirely through $E(a)$. Letting $p,q$ be two points on $\partial E(a)$, and $\gamma \subset E(a)$ the minimal length curve that connects $p$ to $q$ and is contained within $E(a)$. Additionally, let $\gamma'$ be the unique geodesic connecting $p$ to $q$.\footnote{If $p$ and $q$ are far enough separated, there may not be a unique geodesic connecting them. However, the minimal geodesic connecting them will suffice for this proof, and with generic initial data for the matter fields, we expect this degeneracy to be resolved.} If $\gamma \neq \gamma'$, then we could decrease the boundary length of $E(a)$ by instead considering $E(a)'$, defined by replacing $\gamma \to \gamma'$ and filling in the space. This violates the minimality assumption of $E(a)$ unless $\gamma = \gamma'$, and thus $E(a)$ is convex. A similar proof holds in higher dimensions by considering minimal area surfaces instead of curves. 
In higher dimensions, $E(a)$ may no longer be convex,\footnote{We thank Xin-Xiang Ju, Wen-Bin Pan, Ya-Wen Sun and Yuan-Tai Wang for pointing out a subtlety about the convexity of $E(a)$ in more than 2 spatial dimensions.} but its boundary is still the minimal area surface containing $a$.
This result is perhaps most surprising when $a$ is a thin spherical shell, as the region of minimal surface area containing the shell simply fills in the missing sphere in its interior.

In the de Sitter vacuum, however, $E(a)$ will expand to cover all of $\Sigma$, independent of the size of $a$ because the area will always be minimized by covering the whole space, as $\Sigma$ is compact. This implies that the entropy of any subregion in a de Sitter spacetime (or any closed universe) is zero, because the boundary of $E(a)$ is trivial. However, if we put an observer in the complementary patch of $a$, the constraints imposed by the observer may change this result. For example, if we model the observer as an excluded point from $a'$, and provide sufficient entanglement between this observer and the rest of the complementary patch $a'$, then $E(a)$ will minimize the generalized entropy by excluding $a'$, implying a non-zero answer. This qualitatively matches \cite{balasubramanian2023sitter}, and it would be interesting to see if making this observer more precise in our model reproduces the calculations of the entropy of de Sitter spacetimes in \cite{Chandrasekaran:2022cip,balasubramanian2023sitter}.

\subsection{The meaning of our construction: terrestrial holography? } 

Our formula for the entropy recalls the Ryu-Takayanagi (RT) formula in the AdS/CFT correspondence, which says that given a boundary subregion $R$ at a moment of time symmetry, the entropy in the field theory is given by the minimal area surface homologous to $R$ \cite{Ryu_2006}. In fact, in the case that $\Sigma$ is asymptotically AdS and $a$ approaches a boundary subregion, \eqref{eq:BPentropy} precisely matches the RT formula in the $G_N \to 0$ limit. In AdS/CFT, the RT formula is justified via the gravitational path integral \cite{Lewkowycz_2013}, similarly to our approach. It is somewhat surprising that the gravitational path integral, defined on inherently coarse grained degrees of freedom, is sufficient to derive a microscopic entropy formula, which seems like it would need non-perturbative control to compute. For the entropy of black holes, recent work \cite{Penington:2019kki, Balasubramanian:2022gmo} has clarified why the gravitational path integral may actually contain information about the microscopic density of states. We saw in Sec.~\ref{sec:path_int_2} that one way to think about this is that the coarse-grained path integral is still capable of defining the generalized density matrix $\rho_{\mathrm{gen}}$ of the bulk fields which corrects for this coarse-graining. It would be useful to understand whether some similar perspective applies to more general instances of coarse-grained or averaged path integrals computing fine-grained quantities, such as in \cite{Marolf:2020xie,Jafferis:2025vyp,Jafferis:2025yxt,Schlenker:2022dyo,Coleman:1988cy,Giddings:1988cx,Saad:2019lba,Chandra:2022bqq,Saad:2021rcu}.

We know that the RT formula is about more than just entropy: it also determines the bulk information in the gravitating theory that is localized within various subregions of the dual boundary theory. Specifically, given a boundary subregion $R$, and its associated entanglement wedge $E(R)$, any bulk operator with support on $E(R)$ can be reconstructed in the boundary using operators solely living in $R$ \cite{Almheiri_2015}. This property is called entanglement wedge reconstruction and arises entirely from the fact that the RT formula computes entanglement entropy, and the resulting quantum information constraints on holographic encoding of information.\footnote{The RT formula (more generally, the QES formula) determines the entropy of regions, and therefore the relative entropies as well. In \cite{Jafferis_2016}, it was shown that this implies the bulk and boundary modular Hamiltonians agree. Using the the tools of operator algebra reconstruction \cite{Almheiri_2015}, or more generally the Petz map \cite{Chen_2020}, this is sufficient to reconstruct the bulk degrees of freedom in the boundary theory.}
The RT formula satisfies many non-trivial consistency checks, which motivated entanglement wedge reconstruction in the first place, such as no-cloning, nesting, and strong sub-additivity. Does our proposal for the entropy of a bulk subregion pass the same tests?

Bousso and Penington demonstrated \cite{Bousso:2022hlz} that a region $E(a)$ that satisfies the conditions we argued for in Sec.~\ref{sec:subregion} \emph{does} satisfy no-cloning, nesting, and strong sub-additivity. Indeed, this is why they proposed that the generalized entropy of $E(a)$ computes the entropy of the subregion $a$. Their result suggests that the bulk subregion $E(a)$ really is computing an entanglement entropy, as opposed to merely computing a coarse-grained entropy. In AdS/CFT, these entangled degrees of freedom are thought to be those of a CFT living on the spacetime boundary. However, these boundary degrees of freedom can be \emph{defined} through the extrapolate dictionary \cite{Balasubramanian:1998sn,banks1998ads,Balasubramanian:1999ri, Hamilton:2006az,harlow2011operator}. Thus, not only is the bulk defined by the boundary theory, but the boundary theory can be defined by limits of the bulk theory. It is therefore natural to ask if one can view holographic encoding not as a map from the bulk to the boundary, but also as a map from the bulk to other bulk subregions. Indeed, this is the scenario proposed in \cite{Bousso:2022hlz,Bousso:2023sya} and also suggested by our entropy computation.

\begin{figure}
\centering
\includegraphics[]{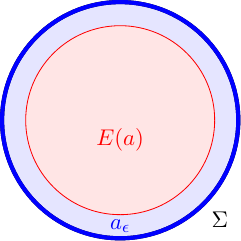}
\caption{A bulk subregion of AdS with \emph{complete} information about the boundary can reconstruct the entire spacetime.}
\label{fig:bulk_band}
\end{figure}

We can further motivate a notion of bulk-to-bulk holography as follows, using the scenarios outlined in \cite{Bousso:2022hlz}. Sticking to AdS spacetimes, where the dual theory is better understood, let $a_\epsilon$ be a spherical band containing the asymptotic boundary and with radius $\rho=\epsilon$ in global coordinates, as in Fig.~\ref{fig:bulk_band}. Full knowledge of the theory of quantum gravity within $a_\epsilon$ includes full knowledge of the asymptotic boundary by the extrapolate dictionary. Because the entanglement wedge of the entire boundary is the entire spacetime, this suggests that in the full theory of gravity, $a_\epsilon$ encodes the full slice $\Sigma$ as well. In fact, this argument generalizes to any bulk region $a_R$ which contains a boundary subregion $R$, i.e. $a_R$ should encode all of $E(R)$ as well, even though it is a bulk region. Furthermore, consider a spacetime containing an evaporating black hole. If the Hawking radiation is collected in an external reservoir, one can show that the interior of the black hole is encoded in the Hawking radiation after the Page time \cite{Marolf_2021,Penington:2019kki,Almheiri:2019hni,Hartman_2020}. Specifically, an ``island'' behind the black hole horizon is holographically encoded in the distant radiation. But now suppose that the Hawking radiation does not escape to the reservoir, but instead is collected at a finite but large distance from the black hole, say in the memory bank of a quantum computer. Performing the same quantum computations as if the radiation were in the reservoir, we expect that the interior is still  encoded within the Hawking radiation, i.e., that ``escaping past infinity'' is not actually crucial to the physics of information recovery through entanglement islands. But because the Hawking radiation is still localized within the bulk, it would appear that we should view the island as an entanglement wedge for the quantum computer, a finite region within the gravitating system. Finally, this idea of a bulk-to-bulk encoding has motivations in tensor network models of quantum gravity \cite{Bousso:2022hlz,Pastawski_2015}. 

\begin{figure}
\centering
\includegraphics{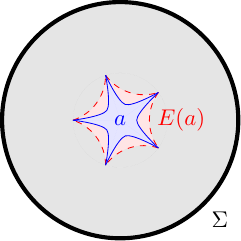}
\caption{A star shaped region $a$ in the AdS vacuum, which should be smoothed out near any kinks. It has arbitrarily small overlap with $b :=\mathcal{B}_\epsilon(\partial E(a))$, an open neighborhood of the corner of $E(a)$. Because $E(a) = E(b)$, these two regions have the same entropy despite being quite distinct geometrically.}
\label{fig:star}
\end{figure}

For more examples, consider the case when $a$ is a star shaped region,\footnote{We thank Chitraang Murdia for this example.} an example of which is shown in Fig.~\ref{fig:star}, smoothing out the sharp kinks in $\partial a$ as needed. Similarly to the example shown in Fig.~\ref{fig:bulkew}, $E(a)$ will fill out the gaps between the ``legs'' of the star. However, consider a second region $b$, which is taken to be thin-but-finite shell with outer boundary $\partial E(a)$. In other words, $b$ is the intersection of $E(a)$ and 
a small open neighborhood of $\partial E(a)$. Then by construction, $E(a)=E(b)$, despite the fact that $a \cap b$ can be arbitrarily small. This suggests that $a$ and $b$ both contain the ``same information'', despite being quite distinct geometrically. Understanding the consistency of these two representations of $E(a)$ will clearly require non-trivial quantum error correcting properties of the map between $a,b$ to $E(a)$. Finally, we argued in Sec.~\ref{sec:applications} that for any closed universe with a moment of time symmetry, any open set $a$ has $E(a) = \Sigma$, which suggests that any open set of a closed universe has complete information about the entire spacetime. This is very similar to what was found in \cite{chakraborty2023holography}, but it is important to note that they obtained this result by studying the Hilbert space associated to a late/early time Cauchy slice, as opposed to the finite time static slice we consider in this paper. Assuming that $E(a)$ defined in our paper is truly analogous to an entanglement wedge, the same result holds in essentially two opposite regimes, providing evidence that this strange feature of closed universes is true for all times.

Another check of our proposal in AdS comes from previous works \cite{Balasubramanian:2013rqa,Balasubramanian_2014,Myers_2014} which have investigated the entropy of a finite region of spacetime by exploiting the purity of the global state $\ket{\Psi}$ and computing the ``differential entropy''  associated to a time band outside the region.
A puzzle in these calculations involved the treatment of non-convex regions \cite{Czech:2015qta,ju2023generalized,ju2023squashed}, as the time bands they construct will always have a convex Rindler horizon.  Our calculation suggests a potential resolution to this puzzle in $2+1$ dimensions. Let $a$ be a non-convex gravitating region in $2+1$d with entanglement wedge $E(a)$. If we assume that the system satisfies complementary recovery, then the complementary degrees of freedom are captured by $E(a)'$, namely the points in $\Sigma$ outside $E(a)$. But $E(a)$ is the convex hull of $a$, and so the domain of dependence $D[E(a)']$ is precisely of the form of the time bands studied in \cite{Balasubramanian_2014,Myers_2014}. This implies that in AdS/CFT, the entropy of $a$ is equal to the entropy of the time band, a requirement for the global state to be pure. If the entropy of $a$ was the area of $a$ and not $E(a)$, this would not have been the case.

Finally, return to the case mentioned in Sec.~\ref{sec:applications} of a thin spherical shell. Our entropy formula suggests that all the degrees of freedom in a compact region of space $a$ live at an open neighborhood of the boundary of $E(a)$. This is reminiscent of the original conception of the holographic principle by 't Hooft and Susskind \cite{tHooft:1993dmi,Susskind_1995}.
These examples strongly motivate the existence of a more ``terrestrial'' formulation of holography, where the dual degrees of freedom are located ``closer to us'' within compact regions of the gravitating spacetime, as opposed to ``the heavens'' at asymptotic infinity.\footnote{Furthermore, the extended corner symmetry group ECS, which played a fundamental role in our computation of the entropy, can be thought of as a finite distance version of the BMS group, which plays a key role in celestial holography.} It would be very interesting to more precisely define this ``dual theory''.

\paragraph{Acknowledgments:} We thank Jon Sorce, Jonah Kudler-Flam, Anthony Speranza, Sam Leutheusser, and Gautam Satishchandran for detailed comments on an early version of this paper.  We are also grateful to Chitraang Murdia, Onkar Parrikar, Bartek Czech, Wayne Weng, Roberto Emparan, Rob Myers, Suvrat Raju, Raphael Bousso, and Rob Leigh for useful conversations and communications. VB is supported by the DOE through DE-SC0013528 and the QuantISED grant DE-SC0020360. CC is supported by the National Science Foundation Graduate Research Fellowship under Grant No.  DGE-2236662.  
VB and CC thank the Yukawa Institute at the University of Kyoto for hospitality during the YITP-T-23-01 Quantum Information, Quantum Matter and Quantum Gravity workshop.

\bibliographystyle{JHEP}
\bibliography{newbiblio}
\end{document}